\documentclass[9pt,preprint]{sigplanconf}
\pdfoutput=1
\usepackage[normalem]{ulem}
\usepackage[T1]{fontenc}
\usepackage{array}
\usepackage{verbatim}
\usepackage{amsmath}
\usepackage{amssymb}
\usepackage{graphicx}
\usepackage{algorithm} 
\usepackage{algorithmic}
\usepackage{authblk}
\makeatletter

\providecommand{\tabularnewline}{\\}

\makeatother

\begin{document}

\title{Managing Hybrid Main Memories with a Page-Utility Driven Performance Model}
\begin{centering}
\authorinfo{Yang Li$^\dagger$,
            Jongmoo Choi$^\ast$,
            Jin Sun$^\dagger$,
            Saugata Ghose$^\dagger$,
            Hui Wang$^\natural$,
            Justin Meza$^\dagger$,
            Jinglei Ren$^\ddagger$,
            Onur Mutlu$^\dagger$}
           {\centerline{$^\dagger$Carnegie Mellon University, \{yangli1, jins\}@andrew.cmu.edu, \{ghose, meza, onur\}@cmu.edu}
            \centerline{$^\ast$Dankook University, choijm@dankook.ac.kr}
            \centerline{$^\natural$Beihang University, hui.wang@jsi.buaa.edu.cn}
            \centerline{$^\ddagger$Tsinghua University, renjl10@mails.tsinghua.edu.cn}}
           {}
\end{centering}                
\date{}

\maketitle

\begin{abstract}
Hybrid memory systems comprised of dynamic random access memory (DRAM)
and non-volatile memory (NVM) have been proposed to exploit both the
capacity advantage of NVM and the latency and dynamic energy advantages
of DRAM. An important problem for such systems is how to place data
between DRAM and NVM to improve system performance.

In this paper, we devise the first mechanism, called UBM (page Utility
Based hybrid Memory management), that systematically estimates the
system performance benefit of placing a page in DRAM versus NVM and
uses this estimate to guide data placement. UBM's estimation method
consists of two major components. First, it estimates how much an
application's stall time can be reduced if the accessed page is placed
in DRAM. To do this, UBM comprehensively considers access frequency,
row buffer locality, and memory level parallelism (MLP) to estimate
the application's stall time reduction. Second, UBM estimates how
much each application's stall time reduction contributes to overall
system performance. Based on this estimation method, UBM can determine
and place the most critical data in DRAM to directly optimize system
performance. Experimental results show that UBM improves system performance
by 14\% on average (and up to 39\%) compared to the best of three state-of-the-art mechanisms
for a large number of data-intensive workloads from the SPEC CPU2006
and Yahoo Cloud Serving Benchmark (YCSB) suites. 
\end{abstract}

\section{Introduction}

Dynamic random access memory (DRAM) has the advantages of relatively
low latency and low dynamic energy, which make it a popular option
for current main memory system designs. However, it is predicted that
DRAM scaling will become increasingly expensive due to increasing
leakage current and manufacturing reliability issues \cite{lbm_2002,ibm_2008,itrs}.
Since data-intensive applications, such as cloud computing and big
data workloads, are becoming widespread, some emerging non-volatile
memory (NVM) technologies (e.g., PCM \cite{Architect_PCM_ISCA2009,Architect_PCM_MicroTopPicks},
STT-RAM \cite{Architect_STTRAM} and ReRAM \cite{ReRam}) have shown
promise for future main memory system designs to meet the increasing
memory capacity demands. 

NVM cells can be more easily manufactured at smaller feature sizes
than DRAM cells, thereby achieving high density and capacity \cite{Architect_PCM_ISCA2009,Architect_PCM_MicroTopPicks,Architect_STTRAM,ReRam,ibm_2008,JSSC2013,IEDM2010,IEDM2009,Qureshi_2009,ISCA2009_Pittsburgh}.
However, NVMs incur high access latency and high dynamic energy consumption,
and some have limited write endurance. In order to address these weaknesses
of NVM, \textit{hybrid memory systems} comprised of both DRAM and
NVM have been proposed to benefit from the large memory capacity of
NVM, while trying to achieve the low latency and low dynamic energy
consumption of DRAM. A key question arising from the use of such hybrid
memory systems is how to manage data placement between DRAM and NVM
to achieve the best of both technologies. In this paper, we aim to
provide a new comprehensive mechanism to optimize overall system performance in hybrid memory
systems.

Most previous proposals on hybrid DRAM-NVM main memory systems either
treat DRAM as a conventional cache \cite{Qureshi_2009} or place data
with high access frequency, high write intensity, and/or low row buffer
locality in DRAM \cite{PDRAM_2009,chop,Zhang_3d_pact2009,Ramos_2011,RowBufferLocality},
while placing the remaining data in NVM. Since the access latency
of NVM is generally higher than that of DRAM (especially for write
requests), placing the frequently accessed data in DRAM can allow
most accesses to benefit from the short access latency of DRAM and
thus improve system performance.

The common characteristic of these previous proposals is that they
consider at most only a few aspects of data characteristics
when constructing a heuristic to guide data placement, without providing a comprehensive model for the performance impact of data placement decisions. These heuristic
metrics can correlate with system performance, but do not directly
capture the system performance benefits of placing different data
in different devices of the hybrid memory system. Therefore, these
proposals can only \textit{indirectly} optimize system performance
and lead to sub-optimal data placement decisions. For example, if
we consider only access frequency \cite{chop} and row buffer locality
to guide data placement \cite{RowBufferLocality}, we may migrate
some data with high access frequency and low row buffer locality from
NVM to DRAM and reduce the latency of accessing this data. However,
it is possible that the memory requests accessing this data are usually
overlapped with other memory requests to NVM from the same application.
As a result, reducing the latency for accessing this data will contribute less to the application's overall stall time reduction, since the concurrent requests to the other data in NVM
may still complete slowly and dominate the stall time of the processor.
Hence, system performance may not benefit, or may benefit less, from
placing data with just high access frequency and low row buffer locality
in DRAM. Therefore, it is desirable to use a comprehensive performance model to directly estimate the performance
benefit of placing a piece of data in DRAM versus NVM, instead of
using only a few metrics as incomplete proxies for performance benefit.

Our goal in this work is to devise a mechanism that directly estimates
the system performance benefit of placing a page in DRAM (with a new, comprehensive performance model), and uses
this estimate to place the performance-critical data in fast memory
(DRAM) to \textit{directly} optimize system performance. To this end,
we define the \textit{utility} of a page as the \textit{system performance
benefit} of placing the page in DRAM versus NVM, and propose a method
to quantify the utility for each page. Based on this utility metric,
we propose a \textit{Page }\textbf{\textit{U}}\textit{tility }\textbf{\textit{B}}\textit{ased
Hybrid }\textbf{\textit{M}}\textit{emory Management} mechanism (UBM)
to improve the system performance of hybrid memory systems. UBM is
a hardware mechanism that aims to identify critical data in the
system, and to place this data in DRAM. UBM's utility metric consists
of two major components. First, it estimates how much an application's
stall time can actually be reduced if the accessed page is placed in fast memory
(DRAM). To do this, UBM comprehensively considers the interaction between the access frequency,
row buffer locality, and memory level parallelism (MLP) of each page
to systematically estimate the application's overall
stall time reduction. Second, UBM estimates how much an application's
stall time reduction contributes to overall system performance (i.e.,
\textit{the sensitivity of system performance to the application's
stall time}). By combining these two components, UBM derives the utility
of each page accessed by any application, and uses these utility estimates
to drive its page placement decisions.

In this paper, we make four main contributions:

1. We propose the first utility metric to quantify the potential system
performance benefit of placing a page in DRAM versus NVM for hybrid
memory systems. The utility metric represents the system performance
benefit as a function of 1) an application's stall time reduction
if the accessed page is placed in DRAM, and 2) the sensitivity of
system performance to each application's stall time.

2. We propose a new performance model that comprehensively considers the access frequency,
row buffer locality, and MLP of a page to systematically estimate an
application's stall time reduction from placing the page in DRAM.
This is the first work that considers MLP in addition to access frequency
and row buffer locality, and models the interaction between them, in page placement decisions.

3. We observe that even when different applications achieve a similar
stall time reduction, the resulting system performance improvement
may be different, which means that system performance exhibits different
sensitivity to the stall time of different applications. We propose
a new method to estimate how much an application's stall time reduction
affects entire system performance, with the goal of prioritizing those applications
that benefit system performance more during page placement.

4. Based on our new performance models (in 2 \& 3) and our new mechanism to estimate them
online, we propose the first page utility based hybrid memory management
mechanism, which selectively places pages that are most beneficial to
overall system performance in DRAM. Our experimental results show that our
proposed mechanism improves system performance by 14\% on average (and up to 39\%) 
compared to the best of three state-of-art mechanisms that we evaluate
(a conventional cache insertion mechanism \cite{Qureshi_2009}, an
access frequency based mechanism \cite{chop,Ramos_2011}, and a row
buffer locality based mechanism \cite{RowBufferLocality}) for a large
number of data-intensive workloads from the SPEC CPU2006 and Yahoo
Cloud Serving Benchmark (YCSB) suites.

\section{Background}

In this section, we introduce the organization and management of DRAM-NVM
hybrid memory systems.

\subsection{DRAM-NVM Hybrid Memory System}

\begin{figure}[tbh]
\centering

\includegraphics[bb=10bp 170bp 680bp 520bp,clip,width=7.5cm]{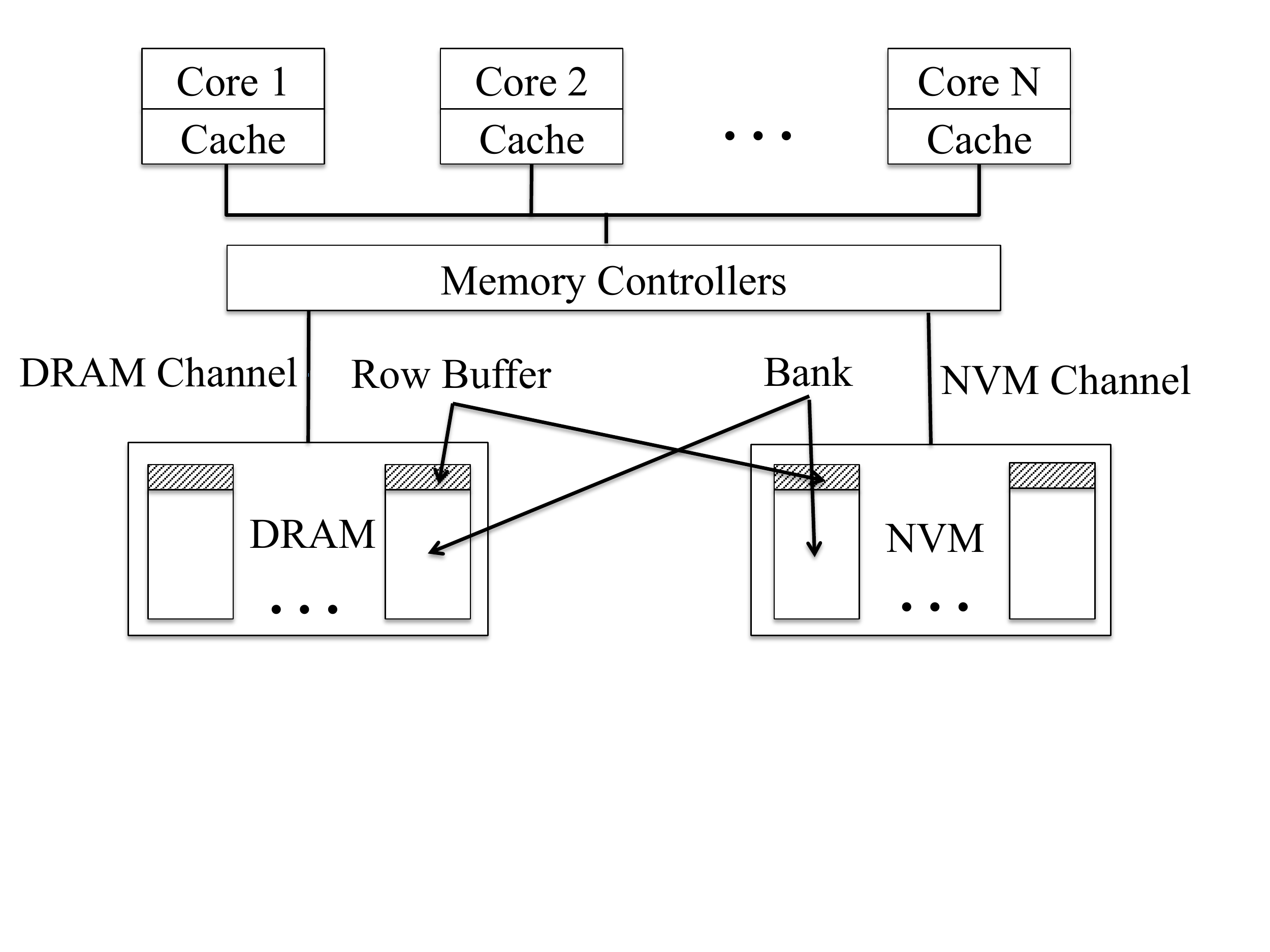}\nocaptionrule

\protect\caption{\label{fig:A-typical-DRAM-NVM}A typical DRAM-NVM hybrid memory system}
\end{figure}

The baseline hybrid memory system in our study is shown as Figure
\ref{fig:A-typical-DRAM-NVM}. In this hybrid memory system, DRAM
and NVM have separate channels to communicate with memory controllers.
When a memory request is issued, memory controllers will determine
whether the request should be sent to DRAM or NVM channel (details
are provided afterwards). The organization of each DRAM/NVM channel
is similar to today's DRAM channel organization. A channel consists
of one or more ranks (omitted in Figure \ref{fig:A-typical-DRAM-NVM}).
Each rank, in turn, consists of multiple banks. Each bank can operate
in parallel, but all banks on a channel share the address and data
bus of the channel.

Each bank has an internal buffer called the row buffer. When data
is accessed from a bank, the entire row containing the data is brought
into the row buffer. Hence, a subsequent access to data from the same
row can be served from the row buffer and need not access the array.
Such an access is called a row buffer hit. If a subsequent access
is to data in a different row, the contents of the row buffer need
to be written back to the row and the new row's contents need to be
brought into the row buffer. Such an access is called a row buffer
miss. A row buffer miss incurs a much higher latency than a row buffer
hit. Previous works on DRAM-NVM memory systems observe that the latency
of a row buffer hit is similar for DRAM and NVM, while the latency
of a row buffer miss is generally much higher in NVM \cite{RowBufferLocality,Architect_PCM_ISCA2009,Architect_PCM_MicroTopPicks}.

An important issue of the hybrid memory system is how to determine
whether a request should be sent to DRAM or NVM. To do this, Qureshi
et al. \cite{Qureshi_2009} proposed to organize DRAM as a cache of
NVM with a 4 KB block size and implement the associated tag store
on chip. The tag store helps memory controllers to determine where
to send the request. In their design, they need a tag store consuming
1 MB SRAM for a hybrid memory system with 1 GB DRAM and 32 GB NVM.
A later proposal \cite{Meza_2012_cal} tried to reduce the on-chip
hardware overhead by storing the tag information in memory (DRAM),
while only leaving the tag information associated with hot pages on
chip. This proposal can significantly reduce the on-chip SRAM consumption
with little performance degradation. 

In our mechanism, we use a similar configuration to these works \cite{Qureshi_2009,Meza_2012_cal},
and organize DRAM as a 16-way set-associative cache with LRU replacement
policy. We assume all data is initially in NVM. Then, instead of migrating
all accessed data used in \cite{Qureshi_2009,Meza_2012_cal}, we selectively
migrate data from NVM to DRAM. When a page is determined to migrate,
we will first check the tag store associated with DRAM to determine
whether the page will evict other data in DRAM. If so, we need to
first evict the victim data from DRAM to NVM, and then migrate the
page from NVM to DRAM. We implement a migration buffer in the memory
controllers to determine the status of the transferred data during
migration. Each cache block of the migrating page is assigned two
bits in the migration buffer to determine where the cache block currently
resides (i.e., in DRAM, NVM or the migration buffer). In this manner,
we can direct incoming memory requests of the migrating page to the
right place. After completing the data movement, corresponding metadata
information in the tag store will be updated. The migration process
between memory devices is fully managed by hardware and transparent
to OS.
\section{\label{sec:Mot}Motivation}

When a page is migrated from NVM to DRAM, the latency of row buffer
misses to that page will decrease.  By combining a page's access frequency
with its row buffer locality, we can determine the total access latency
savings for that page~\cite{RowBufferLocality}.  However, parallelism within
the memory system can prevent a large portion of these latency savings from
translating into performance improvements.
Therefore, in order to estimate the true utility of a page, we
need to 1) estimate the stall time reduction due to the latency reduction,
and 2) estimate how the stall time reduction translates to system
performance improvement (i.e., the sensitivity of overall system performance
to each application's stall time). In this section, we first demonstrate
that we need to \emph{comprehensively} consider access frequency, row buffer
locality, and MLP in order to estimate the stall time reduction, which was not fully
captured in prior metrics~\cite{Zhang_3d_pact2009,chop,Ramos_2011,RowBufferLocality,PDRAM_2009}.
Then, we show that system performance may exhibit different sensitivity
to different applications' stall time.  We take advantage
of this heterogenity between different applications to further optimize the
system performance.

\subsection{\label{sub:Exploiting-MLP-in}Comprehensively Estimating Stall Time
Reduction}

At the first order, an application's stall time reduction depends
on how much the latency for accessing the page can be reduced, as well as how
this latency overlaps with the latencies of other memory requests from
the application. The former can be estimated by considering access
frequency and row buffer locality of the page (i.e., we can just count
the number of row buffer misses to the page, and then estimate the
latency reduction for these memory requests). The latter depends
on the parallelism of the memory requests from an application (MLP). 
MLP measures the number of concurrent outstanding
requests from the same application. In our mechanism, we consider
the MLP for each page and check how many concurrent requests from
the same application typically exist when the page is accessed. If there
are many concurrent requests, the access latency to the page should
overlap with the access latency to other pages, and therefore migrating
the page to DRAM, while it may reduce its access latency, will likely only result in a limited
reduction of the application's stall time. 

\begin{figure}[tbh]
\centering

\includegraphics[bb=10bp 100bp 700bp 430bp,clip,width=8.5cm]{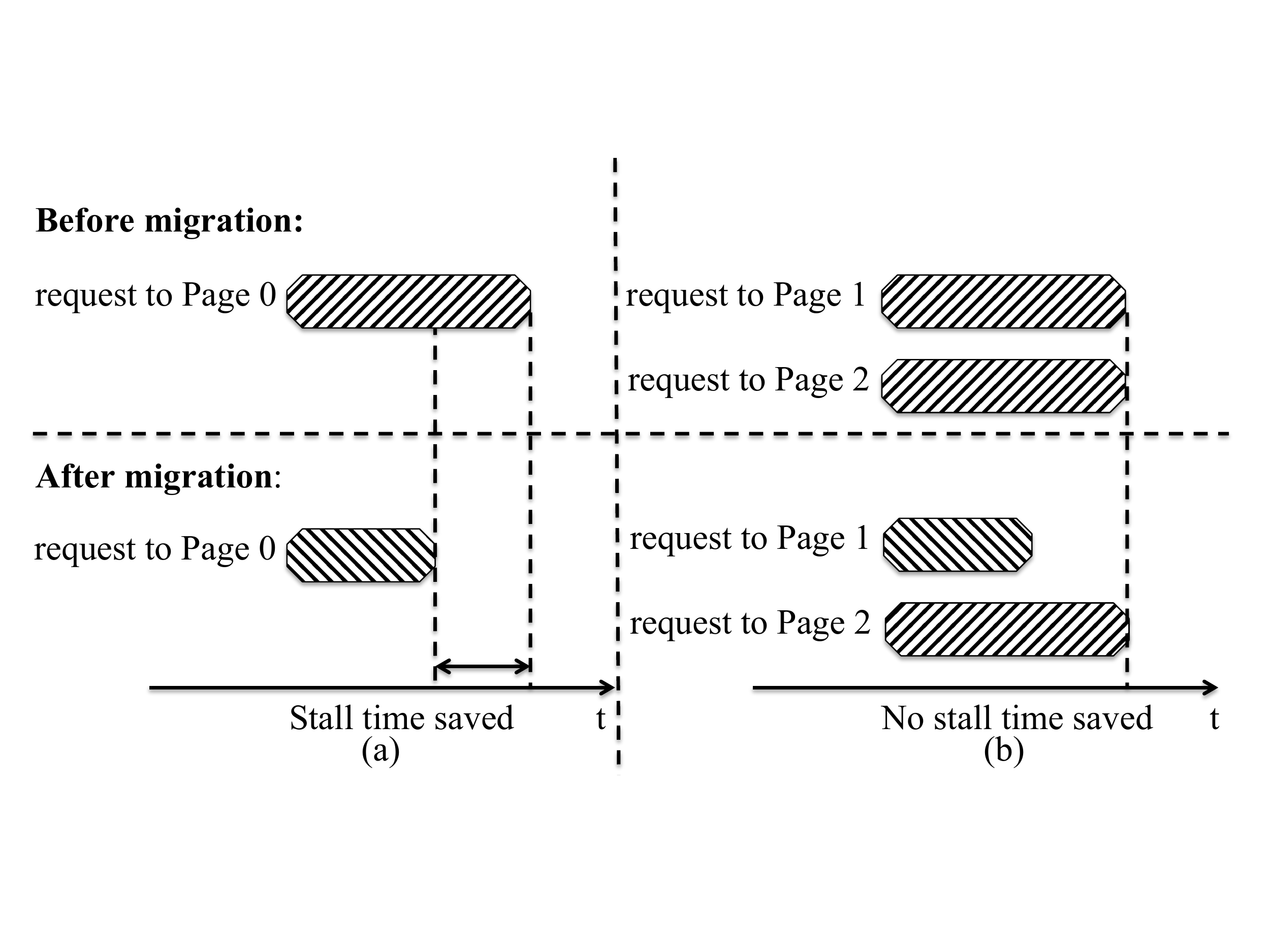}\nocaptionrule

\protect\caption{\label{fig:Conceptual-example-showing}Conceptual example showing
that MLP of a page influences its impact on the corresponding
application's stall time.}
\end{figure}

We can see this effect from the example in Figure \ref{fig:Conceptual-example-showing}.
Pages 0, 1, and 2 all
have the same number of row buffer miss requests.
In our example, requests to Page 0 are usually not overlapped with other
requests from the same application, while requests to Pages 1 and 2
are usually overlapped. We want to see how much the application's
stall time will be reduced if we migrate each of them from NVM to
DRAM.

Suppose we migrate Page 0 to DRAM (Figure \ref{fig:Conceptual-example-showing}a).
As there is no MLP, the request to Page 0 is likely to be stalling at the head
of the processor reorder buffer (ROB).
The requests to Page 0 will complete faster, thereby decreasing the ROB
stall time and being more likely to improve application performance~\cite{clear,runahead,Ghose2013}.
On the other hand, if we migrate Page 1 to DRAM (Figure
\ref{fig:Conceptual-example-showing}b), the requests to Page 1 also
complete faster, but the stall time will not be significantly reduced,
because concurrent requests to Page 2 still maintain the original
access latency and continue to stall at the head of ROB.
Migrating \emph{both} Pages 1 and 2 to DRAM will reduce the stall time,
but that stall time reduction is still roughly the same as that of
migrating Page 0, since the access latency to Page 1 and 2 is overlapped.
Unfortunately, mechanisms that only consider row buffer locality and
access frequency are unable to distinguish between these three pages, and
would suggest migrating Page 1 despite the migration being unhelpful.
Without MLP, we are unable to build a comprehensive model of this behavior.


\begin{figure}[ht]
	\centering
	\begin{minipage}[h]{0.48\linewidth}
		\vspace{-0.5in}
		\centering
		\includegraphics[width=0.98\linewidth]{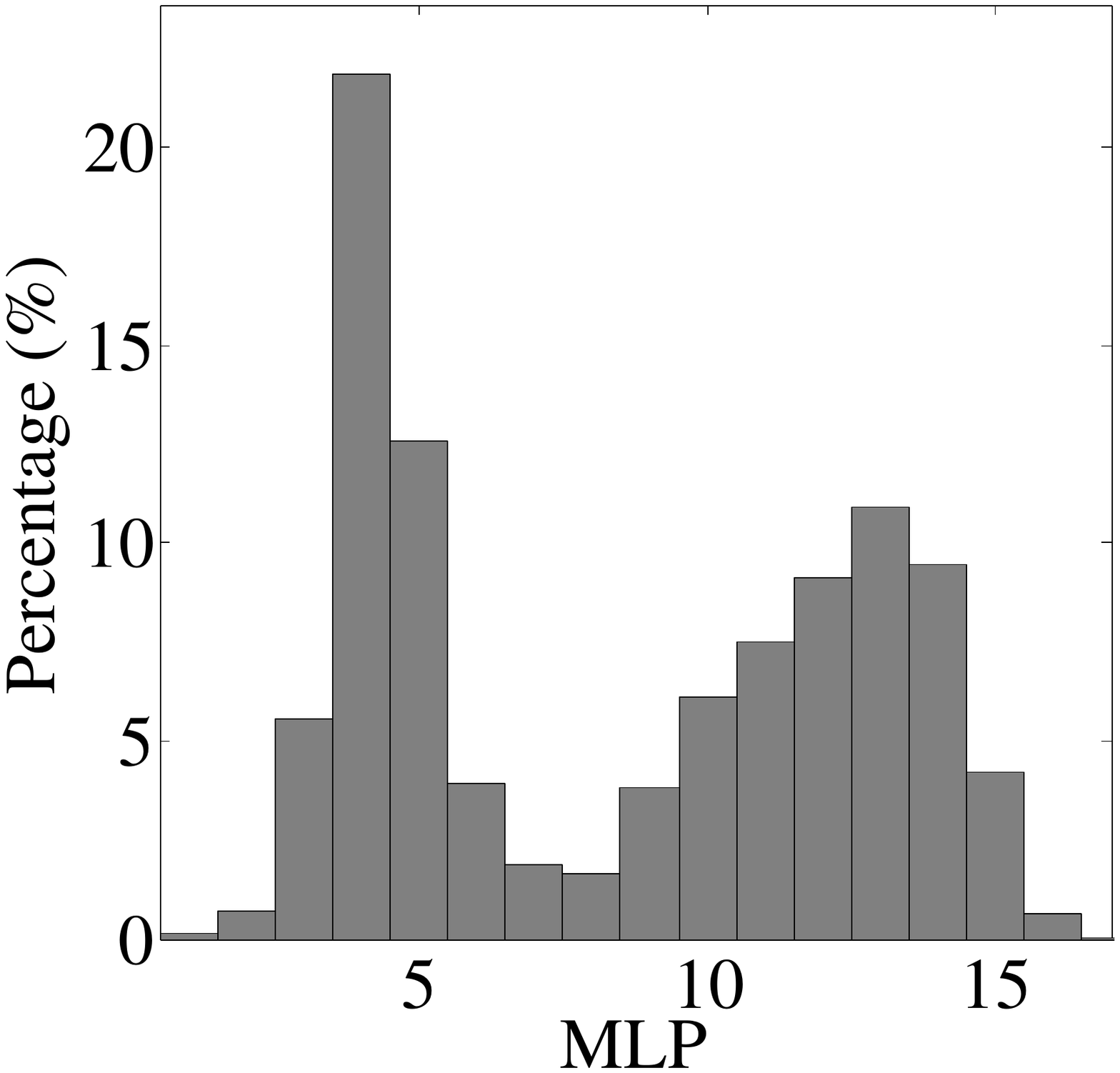}\nocaptionrule
                \vspace{-0.25in}
        \caption{MLP distribution for the pages in xalancbmk.}
        \label{fig:mlpdiversity}
	\end{minipage}
	\begin{minipage}[h]{0.48\linewidth}
		\vspace{-0.4in}
		\centering
		\includegraphics[width=0.98\linewidth]{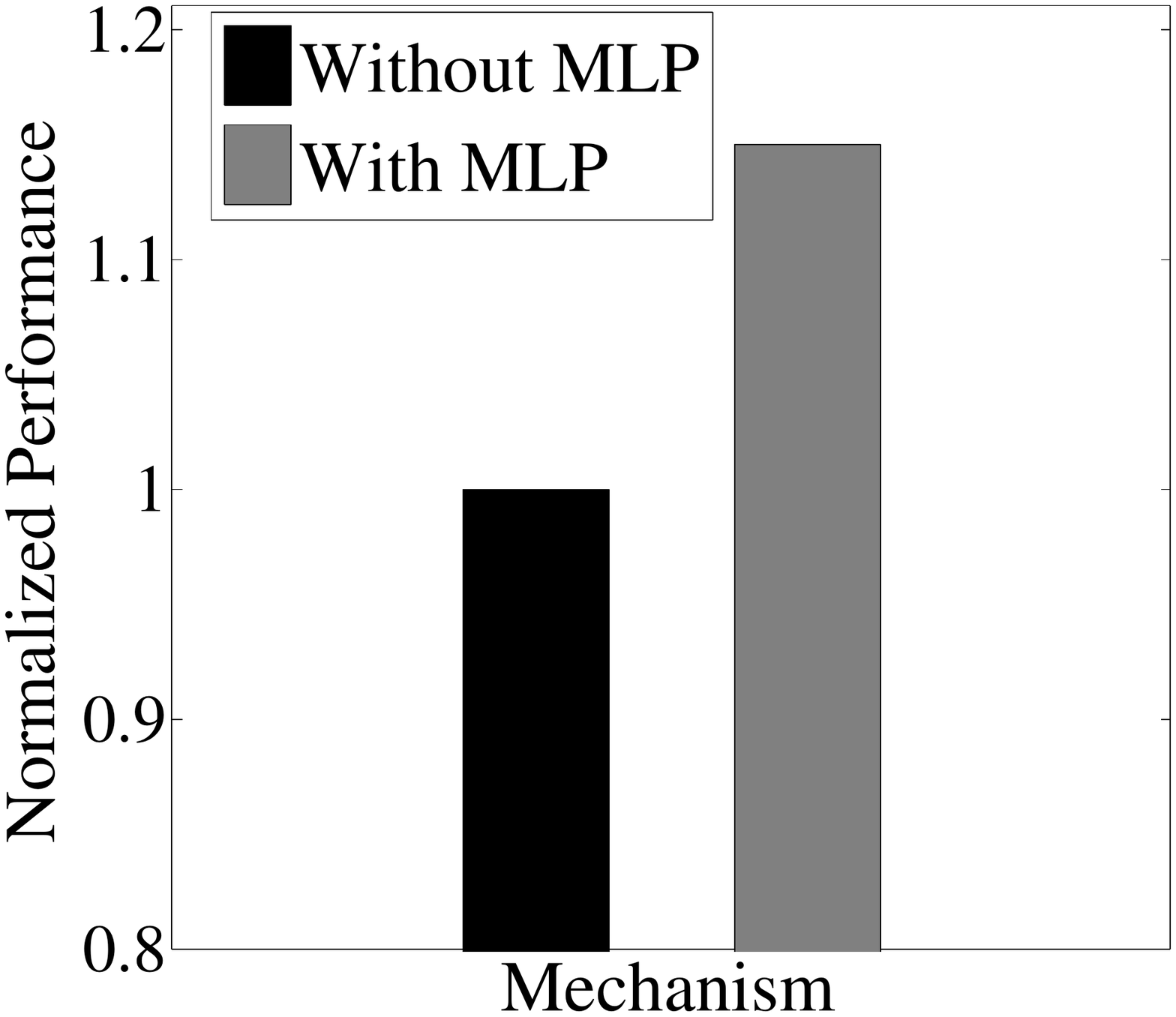}\nocaptionrule
                \vspace{-0.25in}
		\caption{Performance impact of considering MLP on page placement.}
		\label{fig:mlpimpact}
	\end{minipage}
\end{figure}

Figure \ref{fig:mlpdiversity} shows the MLP distribution for the
pages in xalancbmk \cite{SpecCpu}, a representative benchmark. We
can see that different pages within an application may have very different
MLP. Other benchmarks in our evaluation exhibit similar MLP diversity
across pages. Hence, we can take advantage of this diversity
to optimize system performance. Figure
\ref{fig:mlpimpact} shows the performance impact of considering MLP
on page placement for a typical workload.\footnote{This workload consists of soplex, milc, xalancbmk, sphinx3, and astar from SPEC CPU2006, and Workload A, B, and F from the Yahoo Cloud Serving Benchmark (YCSB).}
In this figure, we migrate 64 pages from NVM to DRAM every million
cycles. To select the pages, the mechanism \emph{Without MLP} only
considers access frequency and row buffer locality, and migrates the
pages with the largest number of row buffer miss requests; the
\emph{With MLP} mechanism comprehensively considers access frequency,
row buffer locality and MLP, and migrates the pages with both a large number
of row buffer miss requests and low MLP (which are expected to have the
greatest direct impact on performance). Experimental results show
that incorporating MLP into page placement decisions improves workload 
performance by 15\%. 

In order to quantify the impact of different factors on an application's
stall time, we measure the stall time contribution of each page for
every benchmark in our evaluation. Table 1 shows the correlation
coefficients between the average stall time per page and these
factors (i.e., access frequency, row buffer locality, MLP, and their
combinations).\footnote{For each benchmark, we divide all its pages into
several bins sorted by the values of the factors under consideration. We then
calculate the average stall time per page for each bin. We analyze the
correlation between the average stall time and the factors, and obtain the
correlation coefficient. We report the average correlation coefficient over all
of our benchmarks.} This shows that independently, access frequency, row buffer
locality, and MLP all correlate somewhat with the stall time. However, this
correlation becomes very strong when we comprehensively consider all three
factors together (correlation coefficient = 0.92). We see that the factors
considered in prior work (access frequency and row buffer
locality)~\cite{RowBufferLocality} do not correlate nearly as strongly.
Therefore, we conclude that access frequency, row buffer locality, and MLP are
all indispensable factors to comprehensively model the performance impact of
data placement.

\begin{table}[tbh]
\scriptsize\centering\tabcolsep 0.04in

\begin{tabular}{|c|c|c|c|c|c|c|}
\hline 
 & AF & RBL & MLP & AF+RBL & AF+MLP & AF+RBL+MLP\tabularnewline
\hline 
\hline 
Correlation & 0.74 & 0.59 & 0.54 & 0.76 & 0.86 & 0.92\tabularnewline
\hline 
\end{tabular}

\nocaptionrule

\protect\caption{Absolute Spearman correlation coefficients between the average stall
time per page and different factors (AF: access frequency; RBL: row
buffer locality; MLP: memory level parallelism; the correlation coefficients are between 0 and 1, where 0 = no correlation, and 1 = perfect correlation).}

\end{table}

\subsection{\label{sub:Performance-Sensitivity-to}Sensitivity of System Performance
to Different Applications' Stall Time}

\begin{figure}[tbh]
\centering

\includegraphics[bb=0bp 0bp 1200bp 735bp,clip,width=8.5cm]{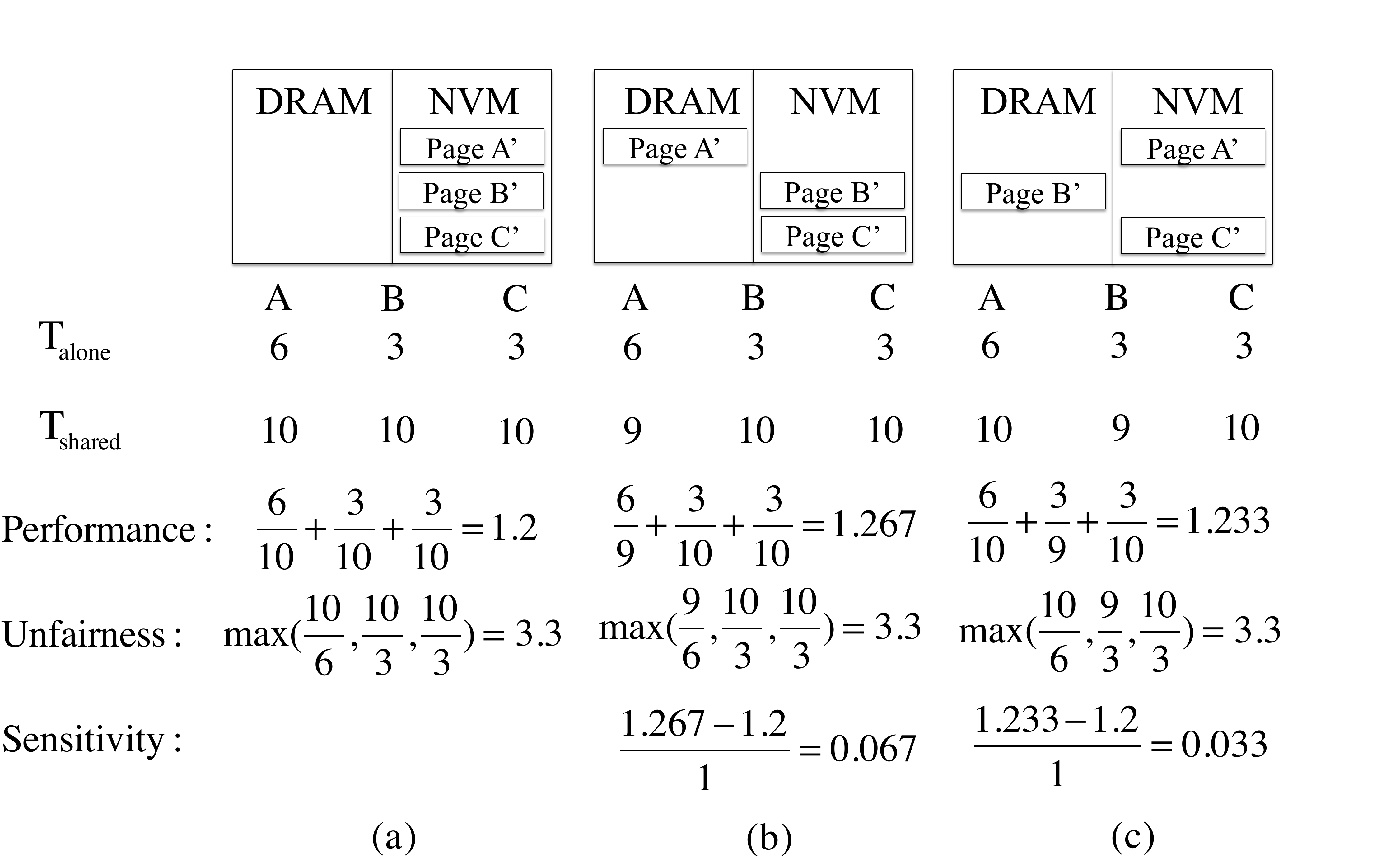}\nocaptionrule

\protect\caption{\label{fig:Conceptual-example-showing-sensitivity}Conceptual example
showing that system performance exhibits different sensitivity to
different applications' stall time (performance metric: weighted speedup
\cite{WeightedSpeedUp}; unfairness metric: maximum slowdown \cite{maxslowdown}).}
\end{figure}

We also observe that system performance may have different sensitivity
to different applications' stall time. Figure \ref{fig:Conceptual-example-showing-sensitivity}
illustrates this observation. In this example, we use weighted speedup
\cite{WeightedSpeedUp} to characterize system performance. Weighted
speedup represents system performance as the sum of the speedup (the execution
time ratio when the application is running alone to that when it is
running with others) of each application. It reflects the system throughput
and quantifies the number of jobs completed per unit of time \cite{metrics}.

In Figure \ref{fig:Conceptual-example-showing-sensitivity}, we show
three applications (A, B, and C) running together. When these applications run alone, their execution times \begin{footnotesize}$T_{alone}$\end{footnotesize}
are 6, 3, and 3, respectively. When run together, their
execution times will increase due to the interference between applications.
Let's consider three pages, A', B', and C', respectively belonging to
applications A, B, and C. Suppose that A', B', and C' are all initially in NVM
(Figure \ref{fig:Conceptual-example-showing-sensitivity}a). Under
this data placement, the execution time \begin{footnotesize}$T_{shared}$\end{footnotesize}
for all three applications is 10. Therefore, the system performance is 1.2
and the unfairness is 3.3. If one of the three pages can be migrated
from NVM to DRAM (which would reduce the stall time of the corresponding application
by 1), we want to see which page, when migrated, would improve
system performance the most. If A' is migrated to DRAM (Figure \ref{fig:Conceptual-example-showing-sensitivity}b),
the execution time of A will decrease by 1, and the system performance
will become 1.267. If B' or C' are migrated to DRAM (Figure \ref{fig:Conceptual-example-showing-sensitivity}c
only shows the case for B' since the case for C' is the same), the
execution time of B or C will decrease by 1 and the system performance
will become 1.233. Therefore, the sensitivity of system performance
to the stall time of A, B, and C is 0.067, 0.033, and
0.033, respectively, and migrating A' improves system performance
the most. (Note that unfairness for the scenarios in Figure
\ref{fig:Conceptual-example-showing-sensitivity}b and \ref{fig:Conceptual-example-showing-sensitivity}c
is the same.) This example implies that we can migrate pages belonging
to the application of which stall time more significantly influences
system performance to DRAM to optimize system performance, while still
maintaining similar fairness guarantee.

\section{Page Utility Based Management (UBM)}

In this section, we introduce the proposed page utility based
hybrid memory management mechanism (UBM). The core of UBM is a utility
metric for each page, which quantifies the expected performance benefit
of migrating the page from NVM to DRAM. The utility metric estimates
the potential stall time reduction due to the page migration, as well as the
sensitivity of system performance to the application's stall time,
and combines these estimations to obtain the integrated system performance
improvement.

\begin{figure}[tbh]
\centering

\includegraphics[bb=35bp 310bp 720bp 510bp,clip,width=7cm]{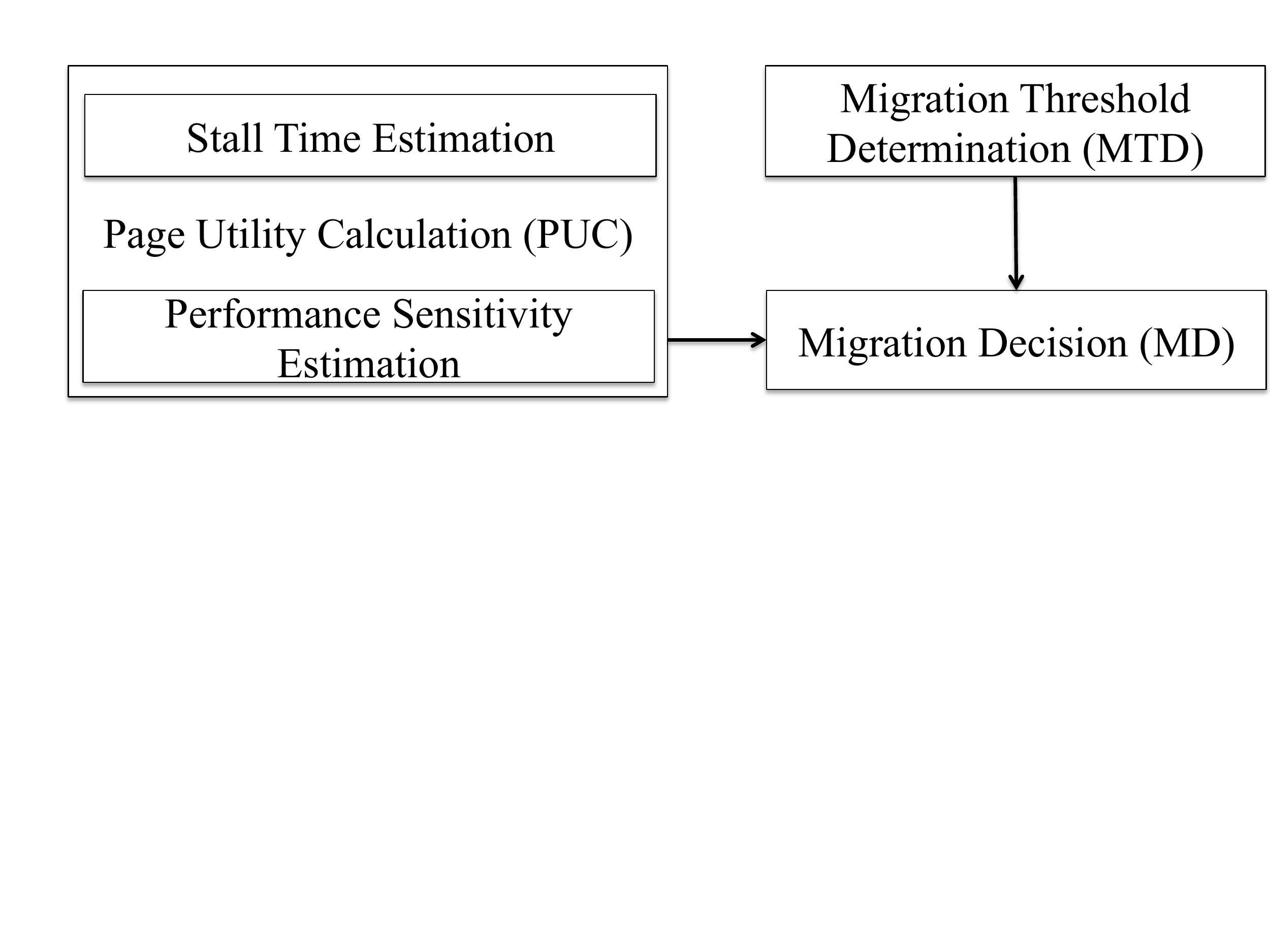}\nocaptionrule

\protect\caption{\label{fig:UBM-Structure}UBM block diagram.}
\end{figure}

Figure~\ref{fig:UBM-Structure} shows an overview of UBM. UBM consists
of three steps: Page Utility Calculation (PUC), Migration Threshold
Determination (MTD), and Migration Decision (MD). Every management
quantum (1 million cycles in our experiments), when an outstanding request completes,
the PUC will update statistical information for the page that was
accessed, and will calculate the utility of that page. This utility
will then be compared with the migration threshold in MD --- if
it exceeds the threshold, the page will be migrated
to DRAM; otherwise, the page will remain in NVM. At the end of each
management quantum, MTD will adjust the migration threshold to maximize
the system performance, based on the statistical information collected
during the quantum.

UBM is a pure hardware mechanism, and is transparent to the OS. In this
section, an ``application'' refers to a hardware thread context.

\vspace{-15pt}

\subsection{Page Utility}
\label{sec:ubm:utility}

We define the utility of a page\begin{scriptsize} ($U$) \end{scriptsize}as the potential change in
system performance if the page is migrated from NVM to DRAM. 
%
As we mentioned in Section~\ref{sec:Mot}, the
utility depends on the stall time reduction of an application, and
the system performance sensitivity to the application. Suppose that
one of the pages of Application $i$ is migrated to DRAM, such that the application
stall time will be reduced by \begin{scriptsize}$\vartriangle Stall\: Time_{i}$\end{scriptsize}. The utility of that page can
be expressed as:
\vspace{-1pt}
\begin{scriptsize}
\begin{equation}
\begin{array}{c}
U=\vartriangle Stall\: Time_{i}\times Sensitivity_{i}\end{array}\label{eq:2}
\vspace{-1pt}
\end{equation}
\end{scriptsize}%
We will elaborate on each of these two components in Section~\ref{sub:Estimation-of-Stall}
and~\ref{sub:Estimation-of-Performance}, respectively. 

\subsubsection{\label{sub:Estimation-of-Stall}Estimating Stall Time Reduction}
\label{sec:ubm:stalltime}

The stall time reduction due to a page migration is dependent on two factors:
(1) the access latency reduction for that page, and (2) the degree to which 
the page's access latency is masked by the access latency of other requests
for the same application.

The degree to which a page's total access latency is reduced can be determined
using a combination of the access frequency and row buffer locality.  If a page
is migrated from NVM to DRAM, the latency of row buffer misses will decrease,
while row buffer hits will still achieve a similar latency.  Therefore, the
expected decrease in access latency is proportional to the total number of row
buffer misses for that page, a function of access frequency and row buffer locality.  We can
estimate this decrease as:
\vspace{-1pt}
\begin{scriptsize}
\begin{equation}
\begin{aligned}\vartriangle Read\: Latency & =\#ReadMiss\times(t_{NVM,read}-t_{DRAM,read})\\
\vartriangle Write\: Latency & =\#WriteMiss\times(t_{NVM,write}-t_{DRAM,write})
\end{aligned}
\label{eq:1}
\vspace{-1pt}
\end{equation}
\end{scriptsize}%
where \begin{scriptsize}$\#ReadMiss$\end{scriptsize} and \begin{scriptsize}$\#WriteMiss$\end{scriptsize} are the number of read and write row
buffer misses, respectively, and \begin{scriptsize}$t_{DRAM,read}$, $t_{DRAM,write}$, 
$t_{NVM,read}$\end{scriptsize}, and \begin{scriptsize}$t_{NVM,write}$\end{scriptsize} are the device-specific read/write
latencies incurred on a row buffer miss.

In order to quantify the degree of access latency masking, we sample the total
number of outstanding memory requests for that same application to model the
``overlap effect.''  Specifically, we define the \emph{MLP ratio} of a page to
be the reciprocal of the outstanding request count.\footnote{We
calculate the MLP ratio separately for reads and writes,
to account for their different behavior in main memory.  While reads are
often serviced as soon as possible (as they can fall along the critical path
of execution), writes will be deferred, and eventually drained in batches.
This allows us to more accurately determine the MLP behavior
affecting each type of request.} Intuitively, if there are fewer outstanding
requests, then there is less memory level parallelism available to overlap the
page's access latency.  As such, we use the reciprocal so that the MLP ratio
represents the fraction of the access latency that will impact the
application's performance.  For a sampling period $t$, for an application with
$N_t$ outstanding requests, the MLP ratio \begin{scriptsize}$MLPRatio_t$\end{scriptsize} will be:
\vspace{-1pt}
\begin{scriptsize}
\begin{equation}
\begin{aligned}MLPRatio{}_{read,t} & \text{=\ensuremath{\frac{1}{N_{read,t}}}} & \quad MLPRatio{}_{write,t}=\frac{1}{N_{write,t}}\end{aligned}
\label{eq:3}
\vspace{-1pt}
\end{equation}
\end{scriptsize}%

Naturally, this MLP ratio will vary throught the quantum, depending on the
transient behavior of the application. As a result, we approximate an average MLP ratio for each page across the entire quantum. Specifically, we use the number of outstanding requests to a page as the weight for the page's MLP ratios at each sampling period, and calculate the weighted average across the whole quantum to represent the average MLP ratios. Using the page's oustanding requests as the weight can better reflect the average extent of how the page's requests are overlapped with other requests from the application. Suppose at sampling period t, the page has \begin{scriptsize}$m_{read,t}$\end{scriptsize}
read and \begin{scriptsize}$m_{write,t}$\end{scriptsize} write
outstanding requests, and the corresponding application has \begin{scriptsize}$N_{read}$\end{scriptsize}
read and \begin{scriptsize}$N_{write}$\end{scriptsize} write
outstanding requests, then further based on Equation \eqref{eq:3},
we can model the page's average MLP ratios during the quantum as:
\vspace{-1pt}
\begin{scriptsize}
\begin{equation}
    \begin{aligned}\overline{MLPRatio{}_{read}} & \text{=\ensuremath{\frac{\underset{t}{\sum}(MLPRatio{}_{read,t}\times m_{read,t})}{\underset{t}{\sum}m_{read,t}}}}=\frac{\underset{t}{\sum}\frac{m_{read,t}}{N_{read,t}}}{\underset{t}{\sum}m_{read,t}}\\
    \overline{MLPRatio{}_{write}} & =\frac{\underset{t}{\sum}(MLPRatio{}_{write,t}\times m_{write,t})}{\underset{t}{\sum}m_{write,t}}=\frac{\underset{t}{\sum}\frac{m_{write,t}}{N_{write,t}}}{\underset{t}{\sum}m_{write,t}}
\end{aligned}
\label{eq:4}
\vspace{-1pt}
\end{equation}
\end{scriptsize}%

We can now combine the latency reduction (Equation~\eqref{eq:1}) and the average
MLP ratio (Equation~\eqref{eq:4}) to determine the stall time reduction for
Application~$i$ as a result of migrating a particular page:
\vspace{-1pt}
\begin{scriptsize}
\begin{equation}
\begin{aligned}\vartriangle Stall\: Time_{i}= & \vartriangle Read\: Latency\times\overline{MLPRatio{}_{read}}\\
 & +p\times\vartriangle Write\: Latency\times\overline{MLPRatio{}_{write}}
\end{aligned}
\label{eq:6}
\vspace{-1pt}
\end{equation}
\end{scriptsize}%
where \begin{scriptsize}$p$\end{scriptsize} represents the probability that the write requests will appear on the
critical path.  Prior work~\cite{ReadWriteRatio} shows that this probability
is dependent on an application's write access pattern, and will generally be
larger if the application has a large number of write requests.  For
simplicity, we choose to set \begin{scriptsize}$p = 1$\end{scriptsize}, though using an iterative online
approach to determine \begin{scriptsize}$p$\end{scriptsize}~\cite{ReadWriteRatio} may yield better performance since it can enhance the accuracy of the stall time estimation.

Equation~\eqref{eq:6} shows that the stall time reduction due to a page migration
from NVM to DRAM can be determined by using a combination of access frequency,
row buffer locality, and MLP for each page.  Intuitively, a high access
frequency and low row buffer locality will increase the number of total row
buffer misses, thus enlarging the benefits of moving to DRAM.  Likewise, poor
MLP, with fewer concurrent outstanding requests, will increase the average
MLP ratio due to low latency masking, and will also increase the benefits from
migration.

\subsubsection{\label{sub:Estimation-of-Performance}Estimating Sensitivity of System
Performance to Different Applications' Stall Time}

For multiprogrammed workloads, we can use the weighted speedup
metric~\cite{WeightedSpeedUp} to characterize system performance.  For each
application, the speedup component of Application $i$ is the ratio of execution
time when running without interference (\begin{scriptsize}$T_{alone,i}$\end{scriptsize}) to that when running
with other applications (\begin{scriptsize}$T_{shared,i}$\end{scriptsize}):
\vspace{-1pt}
\begin{scriptsize}
\begin{equation}
Performance = \sum_i Speedup_{i}=\sum_i \frac{T_{alone,i}}{T_{shared,i}}\label{eq:9}
\end{equation}
\end{scriptsize}
\vspace{-1pt}

When Application \begin{scriptsize}$i$\end{scriptsize} migrates a page to DRAM, the speedup of that application
will improve by $\vartriangle t$:
\vspace{-1pt}
\begin{scriptsize}
\begin{equation}
Speedup_{i}^{'}=\frac{T_{alone,i}}{T_{shared,i}-\vartriangle t}\label{eq:10}
\vspace{-1pt}
\end{equation}
\end{scriptsize}%
Since the stall time reduction due to page migrating is generally much smaller
than the execution time (\begin{scriptsize}$\vartriangle t\ll T_{alone,i},\; T_{shared,i}$\end{scriptsize}), we
can perform a Taylor expansion to find the change in speedup:
\vspace{-1pt}
\begin{scriptsize}
\begin{equation}
\begin{aligned}\vartriangle Speedup_{i} & =Speedup_{i}^{'}-Speedup_{i}=\frac{T_{alone,i}\vartriangle t}{(T_{shared,i}-\vartriangle t)T_{shared,i}}\\
 & \approx\frac{T_{alone,i}}{T_{shared,i}}\cdot\frac{\vartriangle t}{T_{shared,i}}=Speedup_{i}\times\frac{\vartriangle t}{T_{shared,i}}
\end{aligned}
\label{eq:11}
\vspace{-1pt}
\end{equation}
\end{scriptsize}

We defined performance sensitivity in
Section~\ref{sub:Performance-Sensitivity-to} as the measure of how an
application's stall time impacts the overall system performance.  We can
estimate it using Equation~\eqref{eq:11}:
\vspace{-1pt}
\begin{scriptsize}
\begin{equation}
Sensitivity_{i}=\frac{\vartriangle Performance}{\vartriangle Stall\: Time_{i}}=\frac{\vartriangle Speedup_{i}}{\vartriangle t}=\frac{Speedup_{i}}{T_{shared,i}}\label{eq:12}
\vspace{-1pt}
\end{equation}
\end{scriptsize}%

We calculate the performance sensitivity using a quantum-based approach, where
the speedup and execution time obtained in the last quantum are used to
estimate performance sensitivity in the current quantum.  We use the fact that
our quantum is of constant length to transform the sensitivity estimate into an
application speedup estimate.

Equations~\eqref{eq:6} and~\eqref{eq:12} are combined using Equation~\eqref{eq:2} to
give us the overall utility of migrating the page.  Several measurements are
required to obtain this utility calculation, and we will discuss the 
implementation details of these mechanisms in
Sections~\ref{sub:Implementation-Details} and~\ref{sub:Hardware-Cost}.

\subsection{Migration Threshold Determination}

In UBM, once an outstanding request completes, we will recalculate the
utility of its accessed page, and compare this against a migration threshold.
The page will only be migrated from NVM to DRAM if the utility exceeds this
threshold.  A key question is how to determine this migration threshold.

We choose to use a hill climbing based approach
to determine this
threshold dynamically, similar to the policy used by Yoon et
al.~\cite{RowBufferLocality}.  We use the total stall time of all applications
in each quantum to reflect the system performance. At the end of each quantum,
the total stall time will be recalculated. We then compare the current total
stall time with the total stall time from the previous quantum, and determine
whether the previous threshold adjustment yielded a system performance
improvement. If the total stall time decreases in the current quantum (meaning
that the threshold adjustment improved system performance), we continue to
adjust the threshold in the same direction.  Otherwise, since the previous
adjustment degraded performance, we move the threshold in the other direction.

\subsection{\label{sub:Implementation-Details}Implementation Details}

\subsubsection{MLP Ratio Calculation }

We need to calculate MLP ratios for each hot page in NVM, as shown in
Equation~\eqref{eq:4}.  Therefore, we must maintain four temporary counters for
every hot page in the memory controller: \begin{scriptsize}$MLPAcc_{read}$\end{scriptsize} and \begin{scriptsize}$MLPAcc_{write}$\end{scriptsize}
to accumulate the numerator from Equation~\eqref{eq:4}, and \begin{scriptsize}$MLPWeight_{read}$\end{scriptsize}
and \begin{scriptsize}$MLPWeight_{write}$\end{scriptsize} to accumulate the denominator.  For every sampling
period (30 cycles in our experiments), we monitor both the outstanding
read/write requests \begin{scriptsize}$N_{read}$\end{scriptsize} and \begin{scriptsize}$N_{write}$\end{scriptsize} for each application, as well as
the outstanding requests \begin{scriptsize}$m_{read}$\end{scriptsize} and \begin{scriptsize}$m_{write}$\end{scriptsize} for each page, and update
the corresponding counters:
\vspace{-5pt}
\begin{scriptsize}
\begin{equation}
\begin{aligned}MLPAcc_{read}\leftarrow MLPAcc_{read}+\frac{m_{read}}{N_{read}}\;\;\;\quad\quad\;\,\\
MLPAcc_{write}\leftarrow MLPAcc_{write}+\frac{m_{write}}{N_{write}}\qquad\;\:\,\\
MLPWeight_{read}\leftarrow MLPWeight_{read}+m_{read\;\;\;}\\
MLPWeight_{write}\leftarrow MLPWeight_{write}+m_{write}
\end{aligned}
\label{eq:13}
\vspace{-5pt}
\end{equation}
\end{scriptsize}

When all the outstanding requests to a page have completed, the contents
of the page's temporary counters will be added to its corresponding
counters in a \textit{stat store}, and will then be reset. The stat store
is a 32-way set-associative cache with LRU replacement policy residing
in the memory controller.  Each stat store entry corresponds to a page,
consists of six counters that record the number of row buffer misses, the sum
of weighted MLP ratios (\begin{scriptsize}$MLPAcc$\end{scriptsize}), and the sum of weights for the MLP ratios
(\begin{scriptsize}$MLPWeight$\end{scriptsize}) for read/write requests. After an entry's \begin{scriptsize}$MLPAcc$\end{scriptsize} and
\begin{scriptsize}$MLPWeight$\end{scriptsize} counters are updated, the average MLP ratios for the page will be
recalculated with these new values.

\subsubsection{Speedup Estimation}

\begin{table*}[tbh]
\scriptsize\centering\tabcolsep 0.02in%
\begin{tabular}{|>{\raggedright}p{3cm}|>{\raggedright}p{4cm}|>{\raggedright}p{8cm}|l|}
\hline 
Name & Purpose & Structure (sizes in bits in parenthesis) & Size (KB)\tabularnewline
\hline 
\hline 
Stat Store & To track the statistical information of hot pages  & 2048 entries. Each entry consists of read row buffer miss count (8),
write row miss count (8), $MLPAcc_{read}$ (25), $MLPAcc_{write}$
(25), $MLPWeight_{read}$ (15), $MLPWeight_{write}$ (15) and page
number (36) & 33\tabularnewline
\hline 
Counters for outstanding pages in NVM & To record the update of $MLPAcc$ and $MLPWeight$ for pages with
outstanding requests & For each hot page in NVM (96 at most), $MLPAcc_{read}$ (25), $MLPAcc_{write}$
(25), $MLPWeight_{read}$ (15), $MLPWeight_{write}$ (15) and page
number (36) & 1.36\tabularnewline
\hline 
Speedup Estimation & To estimate the speedup of each application & For each application (8 in experiments), $speedup$ (8), $T_{stall}$
(20), $T_{delay}$ (20), $T_{interference}$ (172) \cite{STF_StallTimeFair}  & 0.21\tabularnewline
\hline 
Migration Threshold Determination & To adjust the migration threshold & $Threshold$ (8), $CurrentTotalStallTime$ (23), $PreviousTotalStallTime$
(23), $PreviousAdjustDirection$ (1) & 0.01\tabularnewline
\hline 
\multicolumn{1}{|>{\raggedright}p{3cm}}{Total} & \multicolumn{1}{>{\raggedright}p{4cm}}{} &  & 34.58\tabularnewline
\hline 
\end{tabular}\nocaptionrule

\protect\caption{Main hardware cost of UBM.}
\end{table*}

As mentioned in Section~\ref{sec:ubm:stalltime}, we need to estimate the
speedup of each application in order to determine its sensitivity.  We
modify a mechanism from prior work~\cite{STF_StallTimeFair}, first determining
the additional run time of an application due to contention with other
applications (\begin{scriptsize}$T_{excess}$\end{scriptsize}), and then using this value to calculate the
expected speedup (\begin{scriptsize}$speedup=\frac{T_{alone}}{T_{shared}}=\frac{T_{shared}-T_{excess}}{T_{shared}}=1-\frac{T_{excess}}{T_{shared}}$\end{scriptsize}).

In order to estimate \begin{scriptsize}$T_{excess}$\end{scriptsize}, the method from prior work
systematically considers the interference between applications arising
from bus conflicts, bank conflicts, and row buffer locality changes,
and uses an additional memory request delay (\begin{scriptsize}$T_{interferece}$\end{scriptsize})
caused by the interference from other applications to represent \begin{scriptsize}$T_{excess}$\end{scriptsize}~\cite{STF_StallTimeFair}.
However, applications have the capability to tolerate some memory request
delay, so directly using \begin{scriptsize}$T_{interference}$\end{scriptsize} to model \begin{scriptsize}$T_{excess}$\end{scriptsize} may
overestimate the additional run time. Considering this effect, we
choose to estimate \begin{scriptsize}$T_{excess}$\end{scriptsize} using not only \begin{scriptsize}$T_{interference}$\end{scriptsize}, but
also the total memory request delay (\begin{scriptsize}$T_{delay}$\end{scriptsize})
and the processor stall time (\begin{scriptsize}$T_{stall}$\end{scriptsize}).

The intuition behind our method is that the processor stall time is
caused by the memory request delay, both from the application itself
and from the interference of others. Therefore, we can model the additional
run time as \begin{scriptsize}$T_{excess}=T_{stall}\times\frac{T_{interference}}{T_{delay}}$\end{scriptsize},
using the proportion of interference-caused delay to total delay.
After obtaining the modified additional run time, we use the same approach
as the original method to estimate the speedup. Over the original method,
our modifications only need to add hardware to measure the processor stall time
(\begin{scriptsize}$T_{stall}$\end{scriptsize}) and the length
of the period when the application has outstanding memory requests (\begin{scriptsize}$T_{delay}$\end{scriptsize}).

\subsubsection{Utility Calculation for Shared Pages}

For pages shared by multiple applications, we can use separate
entries in the stat store to record the statistical information of
the page with respect to each application. We can use
our previous method to calculate the page utility for each application,
and then add them together to obtain the aggregate utility for the
page. The insight is that the total system performance improvement
is just the sum of the performance improvement of each application. Therefore,
summing up the page utility for each application (i.e., its performance
improvement) should reflect the system performance
improvement.

\subsection{\label{sub:Hardware-Cost}Hardware Cost}

Table 2 describes the main hardware costs for UBM. The largest
component is the stat store.  We use a 2048-entry store (i.e., 32-way
64-set-associative cache), as it leads to negligible performance degradation
compared with an unlimited stat store. The main hardware
cost of UBM is 34.58KB,\footnote{This does not include the hardware used to determine whether a page resides in DRAM or NVM, as it is required by most hybrid memory management mechanisms, and the implementation of UBM is orthogonal to the implementation of this structure.} which is 1.7\% of L2 cache size. 

UBM also requires circuitry to calculate the MLP ratios.
For each hot page in NVM (96 at most; limited by the NVM read request
queue size and write buffer), we need to perform 4 25-bit additions and 2 fast
divisions every 30~cycles (Equation~\eqref{eq:13}).  We achieve this by
pipelining the logic, and making it 3-way superscalar.
We can implement fast division using a $32\times32$ ROM table that
contains the precomputed results of the division, since both the
numerator and denominator of the division are limited by the MSHR size
of the last-level cache.  As each quotient is 10 bits wide, the total
size of the ROM table will be 1.25KB.

\subsection{\label{sub:Criticality}Comparison with Criticality}
UBM uses data characteristics and application characteristics to approximate the \emph{actual} system performance benefits of placing a page in DRAM, instead of directly measuring the time that the page's load requests stall the processor pipeline to guide data placement (i.e., load criticality~\cite{Ghose2013}). This is because the latter can often mistakenly attribute those stalls caused by writes. Though store instructions commit in the processor before the memory system completes the write operations, these writes can still stall processor progress during a write queue drain. A subsequent load to main memory can stall until the drain finishes (in this case, migrating the page being read would have little impact on this stall time). However, since the writes have already been committed, load criticality is unable to properly attribute the stall to the store operations (oftentimes incorrectly attributing the stalling to the load). This would in turn not migrate the write page, which could be especially costly in NVM due to its longer write latencies. Unlike criticality, UBM can correctly attribute the stall time to the write drain (which it observes directly in the memory controller), and is able to migrate the stall-inducing write pages to improve system performance.

\section{Evaluation Methodology}

\subsection{System Configuration}

We evaluate the proposed UBM mechanism using a cycle-accurate in-house
x86 multicore simulator, whose front end is based on
Pin \cite{pin}. The simulator models the memory system in detail.
In this simulator, the page migrations between DRAM and NVM are modeled
as additional read/write memory requests to these memory devices.
The latency for determining whether a page resides in DRAM or NVM
is modeled as 6 cycles. Table \ref{tab:Baseline-system-parameters} summarizes
the major baseline system parameters in our evaluation, including DRAM and
NVM timing and energy parameters~\cite{Micron,justin_report,Architect_PCM_ISCA2009}.
We also vary the DRAM size and NVM timing parameters
for our sensitivity studies.

\begin{table}[tb]
\scriptsize\centering\tabcolsep 0.02in

\begin{tabular}{|>{\raggedright}p{17mm}|>{\raggedright}p{65mm}|}
\hline 
Processor & 8 cores, 2.67GHz, 3 wide issue, 128-entry instruction window\tabularnewline
\hline 
\hline 
L1 cache & 32KB per core, 4-way, 64B cache block\tabularnewline
\hline 
\hline 
L2 cache & 256KB per core, 8-way, 32 MSHR entry per core, 64B cache block\tabularnewline
\hline 
\hline 
DRAM Memory Controller & 64 bit channel, 64-entry read request queue, 32-entry write buffer, FR-FCFS
scheduling policy \cite{FRFCFS1,FRFCFS2}\tabularnewline
\hline 
\hline 
NVM Memory Controller & 64 bit channel, 64-entry read request queue, 32-entry write buffer, FR-FCFS
scheduling policy\tabularnewline
\hline 
\hline 
DRAM memory system & 512MB, 1 rank (8 banks), 8 KB page size, $t_{CLK}$=1.875ns, $t_{CL}$=15ns,
$t_{RCD}$=15ns, $t_{RP}$=15ns, $t_{WR}$=15ns, array read (write)
energy = 1.17 (0.39) pJ/bit, row buffer read (write) energy = 0.93
(1.02) pJ/bit, standby power = 21 uW/bit \tabularnewline
\hline 
\hline 
NVM memory system & 16GB, 1 rank (8 banks), 8 KB page size, $t_{CLK}$=1.875ns, $t_{CL}$=15ns,
$t_{RCD}$=67.5ns, $t_{RP}$=15ns, $t_{WR}$=180ns, array read (write)
energy = 2.47 (16.82) pJ/bit, row buffer read (write) energy = 0.93
(1.02) pJ/bit, standby power = 21 uW/bit \tabularnewline
\hline 
\end{tabular}

\protect\caption{\label{tab:Baseline-system-parameters}Baseline system parameters.}
\end{table}

\begin{table}[tbh]
\scriptsize

\tabcolsep 0.02in

\centering

\begin{tabular}{|c|c|c|c|c|c|c|c|c|}
\hline 
Benchmark & MPKI & Class & Benchmark & MPKI & Class & Benchmark & MPKI & Class\tabularnewline
\hline 
\hline 
mcf & 100.7 & I & lbm & 52.1 & I & soplex & 45.5 & I\tabularnewline
\hline 
milc & 33.2 & I & omnetpp & 31.9 & I & xalancbmk & 27.4 & I\tabularnewline
\hline 
libquantum & 26.8 & I & GemsFDTD & 15.2 & I & sphinx3 & 13.7 & I\tabularnewline
\hline 
leslie3d & 12.1 & I & bzip2 & 8.93 & I & zeusmp & 7.36 & I\tabularnewline
\hline 
astar & 4.72 & I & YCSB A & 4.22 & I & YCSB F & 3.68 & I\tabularnewline
\hline 
YCSB B & 3.63 & I & tonto & 3.42 & N & YCSB E & 3.38 & N\tabularnewline
\hline 
YCSB D & 3.23 & N & YCSB C & 3.18 & N & h264 & 2.40 & N\tabularnewline
\hline 
perlbench & 1.68 & N & wrf & 0.51 & N & sjeng & 0.49 & N\tabularnewline
\hline 
namd & 0.26 & N & bwaves & 0.20 & N & gobmk & 0.19 & N\tabularnewline
\hline 
gamess & 0.10 & N & povray & 0.07 & N & calculix & 0.02 & N\tabularnewline
\hline 
\end{tabular}

\protect\caption{\label{tab:Characteristics-of-30}Characteristics of 30 SPEC CPU2006
and YCSB benchmarks (I: memory-intensive class; N: non-memory-intensive
class).}
\end{table}

\subsection{\label{sub:WorkloadsSection}Workloads}

We use 30 benchmarks chosen from SPEC CPU2006~\cite{SpecCpu} and the
Yahoo Cloud Serving Benchmark (YCSB) suite~\cite{YCSB}. Each benchmark
was warmed up for 500 million instructions, and then executed for another 500 million
instructions. The warm-up phase is long enough to guarantee that the DRAM hit rate reaches relatively steady state by the end of the phase.\footnote{Note that in steady state, the DRAM may not be full, as some mechanisms take advantage of the separate NVM memory channel to perform request load balancing.  If the entire working set were placed in DRAM, the extra contention on the DRAM memory channel may hurt performance so much that it undoes the benefits of caching, while in the meantime the independent NVM memory channel remains idle, wasting available bandwidth.}

We classify the benchmarks as memory-intensive or non-memory-intensive
based on their last level cache misses per 1K instructions
(MPKI) when running alone. Table~\ref{tab:Characteristics-of-30}
shows the benchmark characterizations. In our experiments, a
workload is grouped using 8 benchmarks, and the workload intensity
is calculated based on the proportion of memory-intensive benchmarks
to total benchmarks. For example, a workload has 75\% intensity
if it consists of 6 memory-intensive benchmarks and 2
non-memory-intensive benchmarks. Using this approach, we generate
40 workloads, which exhibit
0\%, 25\%, 50\%, 75\%, or 100\% workload intensity.

\subsection{Metrics}

We use weighted speedup (\begin{scriptsize}$Wspeedup$\end{scriptsize})~\cite{WeightedSpeedUp}
as the main metric to evaluate the system performance. Weighted speedup
reflects the system throughput, and is suitable for system-oriented
performance quantification~\cite{metrics}. We also provide results
for harmonic speedup (\begin{scriptsize}$Hspeedup$\end{scriptsize})~\cite{harmonic_speedup}, which
reflects the average turnaround time, and is suitable for user-oriented
performance quantification \cite{metrics}. We use maximum
slowdown~\cite{maxslowdown} to evaluate unfairness. These metrics are
shown below. \begin{scriptsize}$N$\end{scriptsize} is the number of cores; \begin{scriptsize}$IPC_{alone,i}$\end{scriptsize} and \begin{scriptsize}$IPC_{shared,i}$\end{scriptsize}
are the system throughput when Application \begin{scriptsize}$i$\end{scriptsize} is running alone and
running with other applications, respectively.

\vspace{-5pt}
\begin{scriptsize}

\[
Wspeedup=\sum_{i=0}^{N-1}\frac{IPC_{shared,i}}{IPC_{alone,i}}\quad Hspeedup=\frac{N}{\sum_{i=0}^{N-1}\frac{IPC_{alone,i}}{IPC_{shared,i}}}
\]

\[
Unfairness=\underset{}{max}\left(\frac{IPC_{alone,i}}{IPC_{shared,i}}\right)
\vspace{-5pt}
\]
\end{scriptsize}

\section{Experimental Results}

We evaluate our proposed UBM mechanism over a variety of system configurations,
ranging over several DRAM sizes and NVM access latencies.  Throughout our
evaluation, we compare UBM against four other mechanisms:
\begin{itemize}
\item \emph{ALL}: a conventional cache insertion mechanism. This mechanism treats
DRAM as a cache to NVM, and inserts all data accessed in NVM into
DRAM using the LRU replacement policy. This is similar
to the proposal by Qureshi et al.~\cite{Qureshi_2009}.
\item \emph{FREQ}: an access frequency based mechanism. This mechanism migrates pages
with high access frequency to DRAM. It is similar to two proposals that try to improve the temporal locality
in DRAM and reduce the number of accesses to NVM~\cite{chop, Ramos_2011}.
\item \emph{RBLA}: a row buffer locality based mechanism~\cite{RowBufferLocality}.
This mechanism migrates pages which have experienced a large number
of NVM row buffer misses to DRAM. The intuition is that only the latency
of row buffer miss requests can be reduced when the page is migrated
to DRAM.
\item \emph{UBM-ST}: a stall time reduction based mechanism (also proposed in this paper). This is a simplified version of
UBM, with a utility metric that only considers the stall time reduction
but neglects the sensitivity of system performance to the application's
stall time\textit{.} UBM-ST will migrate a page from NVM to DRAM if
the estimated stall time reduction is large. This mechanism
can be considered to be a part of UBM, and helps us quantify the significance
of each component of our complete proposed page utility metric.
\end{itemize}

\subsection{\label{sub:EvaluationBaseline}Evaluation for Baseline System Configuration}

\begin{figure*}[ht]
	\centering
	\begin{minipage}[h]{0.32\linewidth}
		\vspace{-0.1in}
		\centering
		\includegraphics[width=0.98\linewidth]{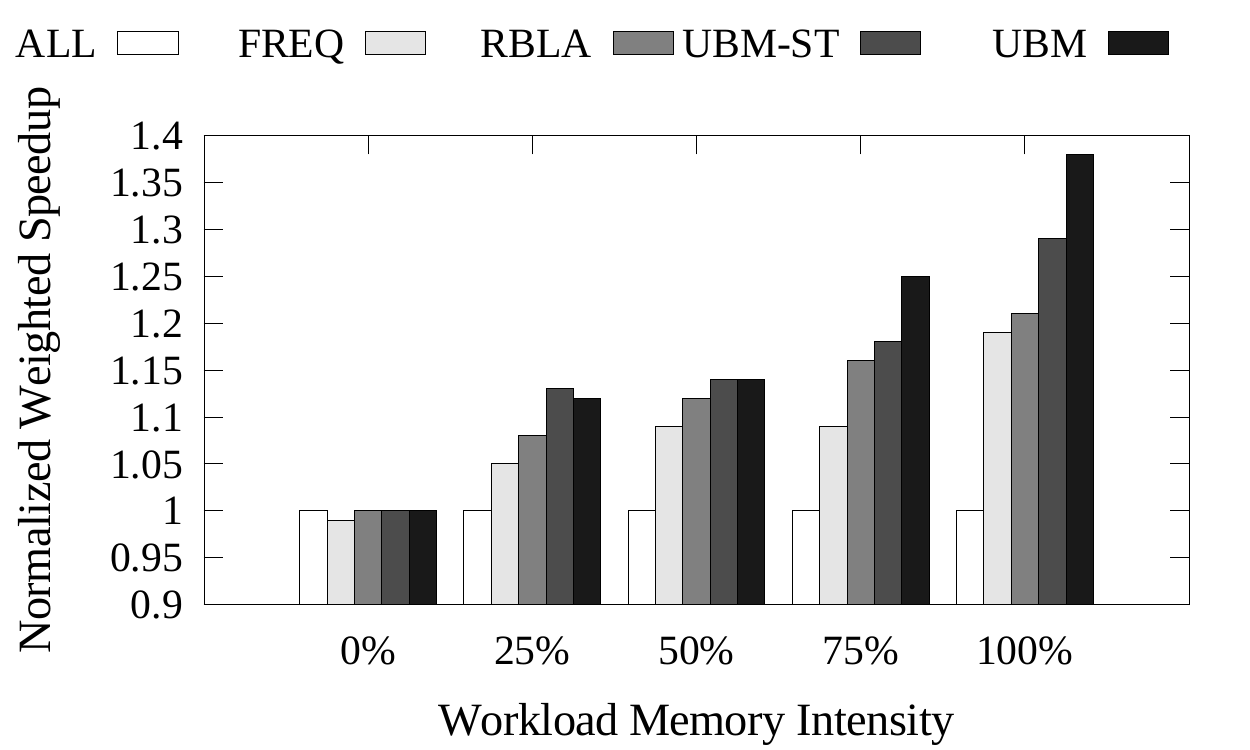}\nocaptionrule
        \caption{Normalized weighted speedup for baseline configuration.}
        \label{fig:Performance-for-baseline}
	\end{minipage}
	\begin{minipage}[h]{0.32\linewidth}
		\vspace{-0.1in}
		\centering
		\includegraphics[width=0.98\linewidth]{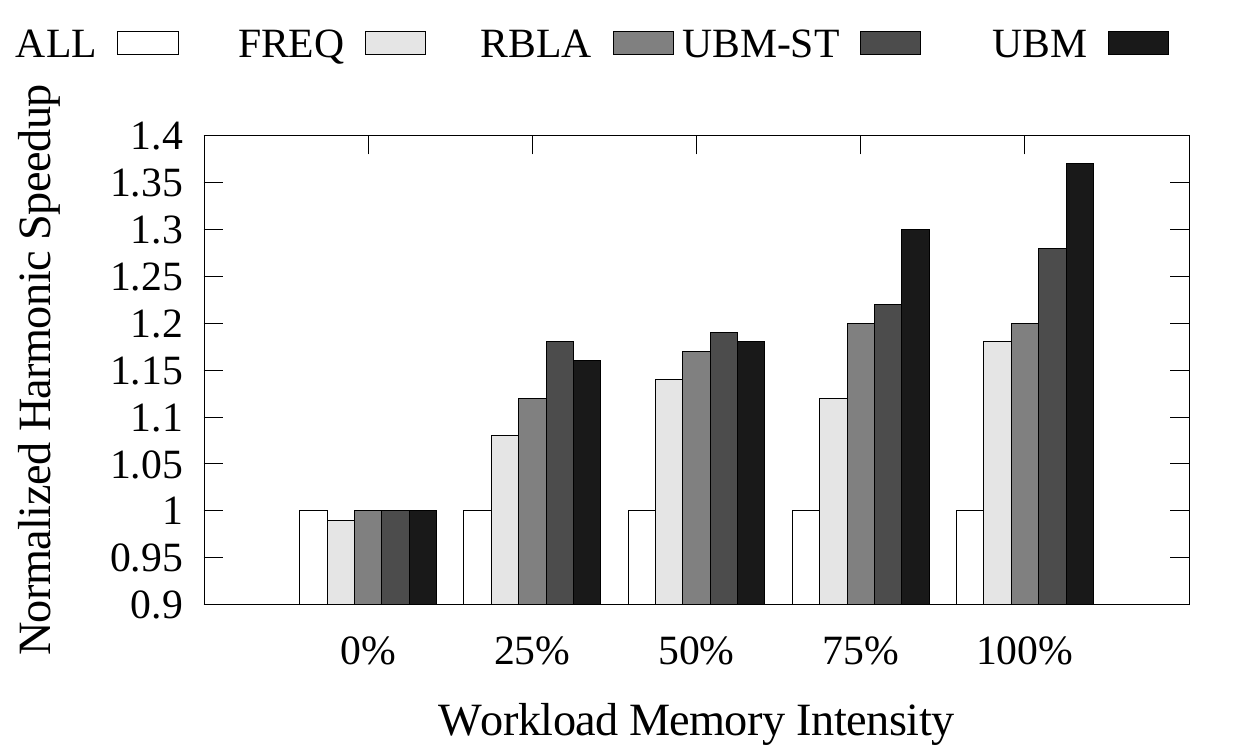}\nocaptionrule
		\caption{Normalized harmonic speedup for baseline configuration.}
		\label{fig:Normalized-harmonic-speedup}
	\end{minipage}
	\begin{minipage}[h]{0.32\linewidth}
		\vspace{-0.1in}
		\centering
		\includegraphics[width=0.98\linewidth]{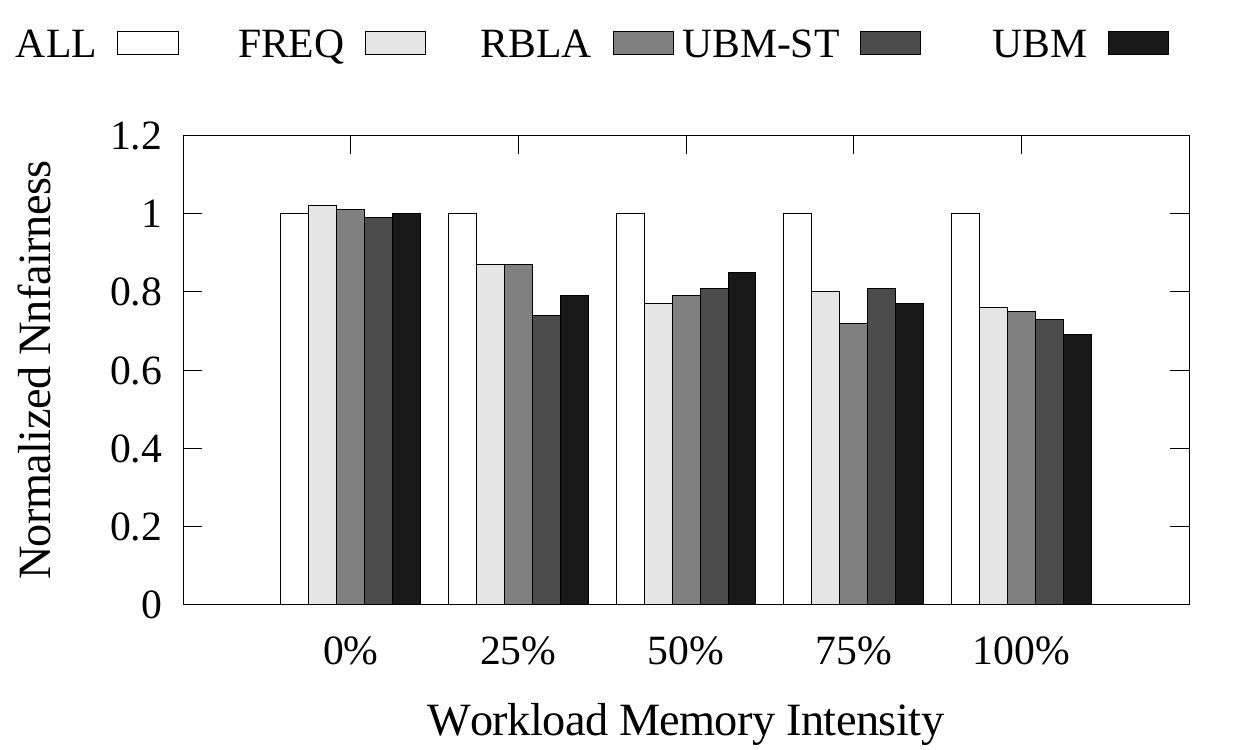}\nocaptionrule
		\caption{Normalized unfairness for baseline configuration.}
		\label{fig:Fairness-for-baseline}
	\end{minipage}

\end{figure*}

\begin{figure*}[ht]
	\centering
	\begin{minipage}[h]{0.32\linewidth}
		\vspace{-0.1in}
		\centering
		\includegraphics[width=0.98\linewidth]{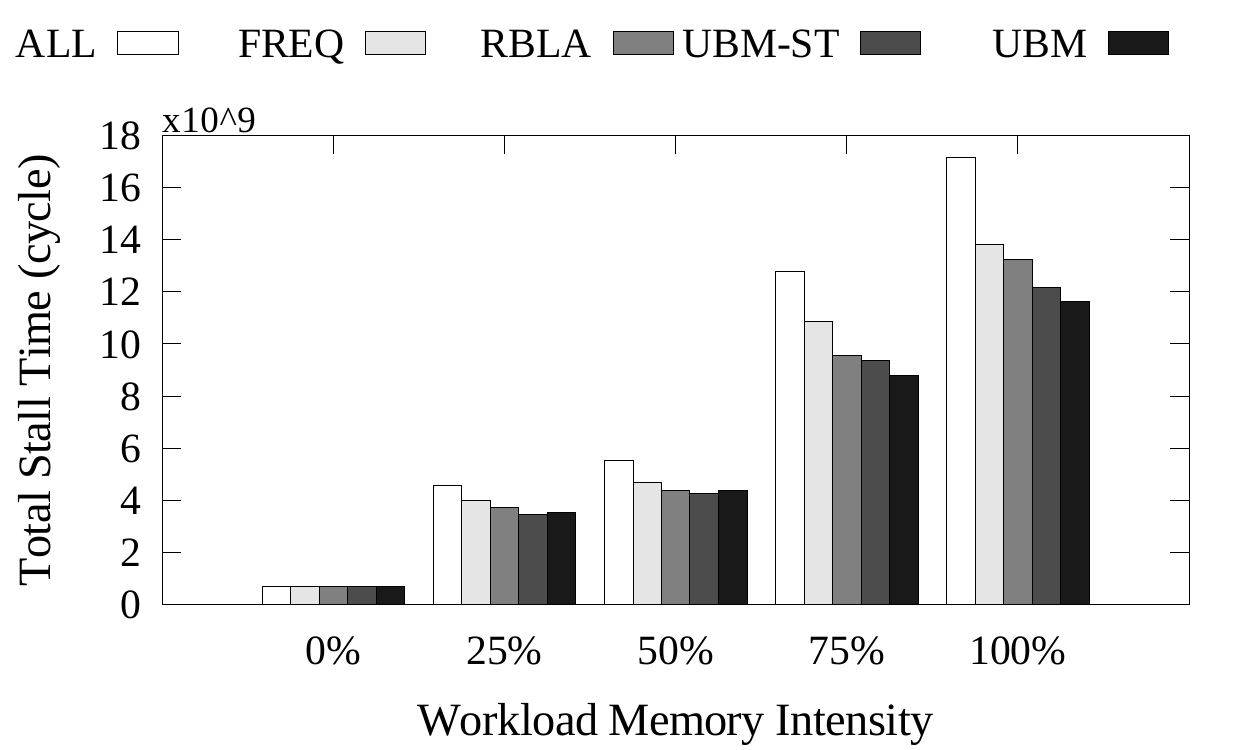}\nocaptionrule
		\caption{Total stall time for baseline configuration.}
		\label{fig:Total-stall-time}
	\end{minipage}
	\centering
	\begin{minipage}[h]{0.32\linewidth}
		\vspace{-0.1in}
		\centering
		\includegraphics[width=0.98\linewidth]{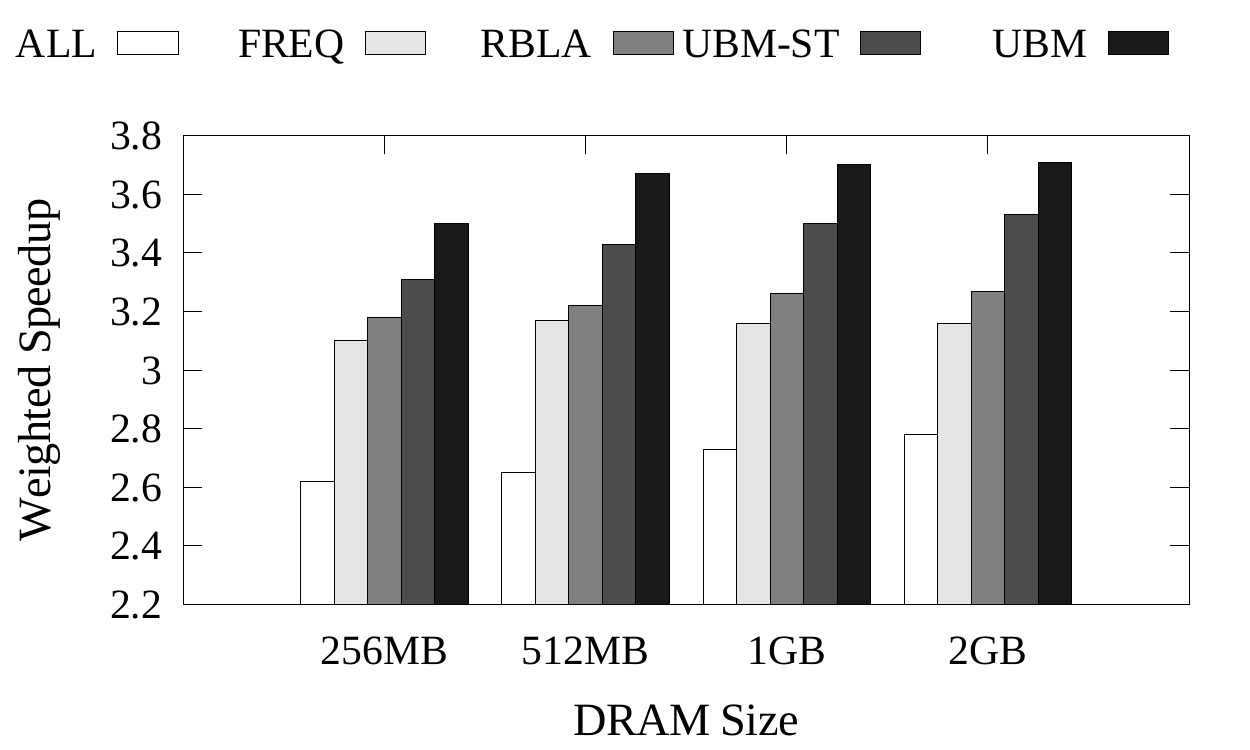}\nocaptionrule
		\caption{Weighted speedup for various DRAM sizes.}
		\label{fig:Performance-for-configurations}
	\end{minipage}
	\centering
	\begin{minipage}[h]{0.32\linewidth}
		\vspace{-0.1in}
		\centering
		\includegraphics[width=0.98\linewidth]{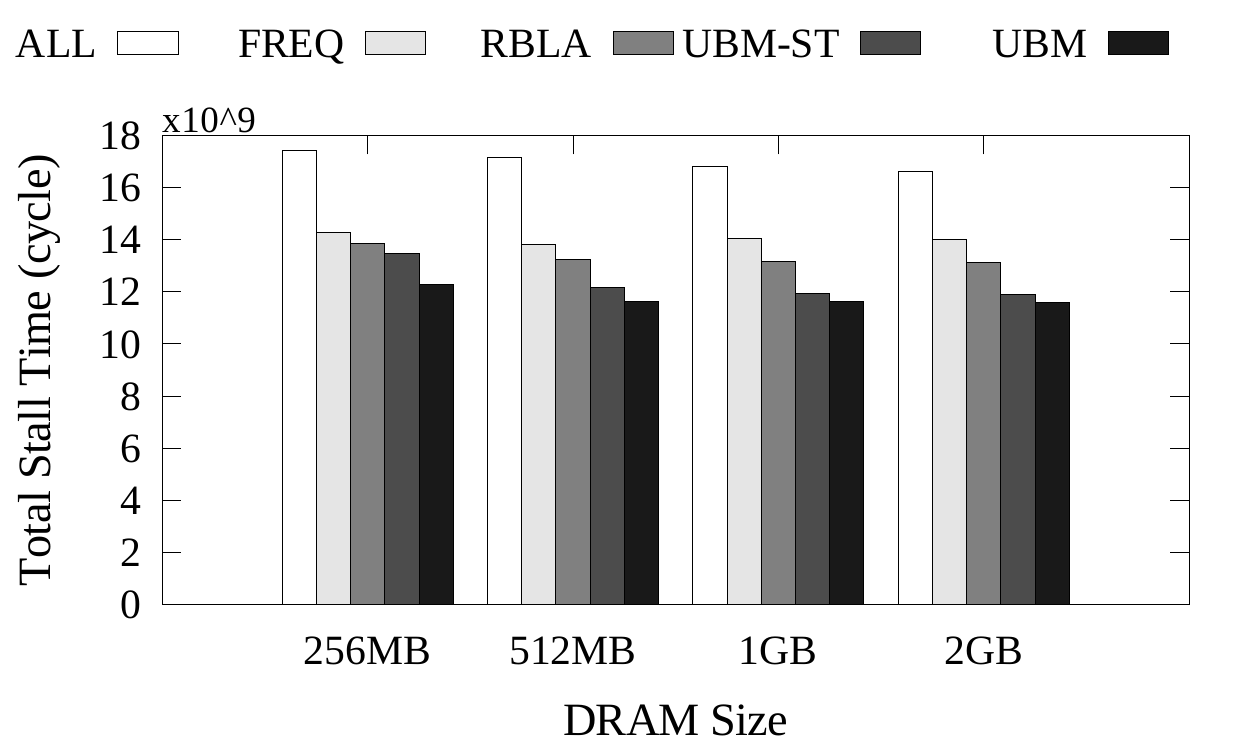}\nocaptionrule
		\caption{Total stall time for various DRAM sizes.}
		\label{fig:Total-stall-time-1}
	\end{minipage}
\end{figure*}

Figure~\ref{fig:Performance-for-baseline} shows the normalized weighted
speedup of these five mechanisms on the baseline system configuration.
We can see that UBM-ST outperforms the best previous proposal, RBLA,
in all workload categories where the memory intensity is larger than
0. For the most memory-intensive category, UBM-ST achieves a 7\% average performance
improvement over RBLA. UBM-ST not only considers the latency of each
individual request (as FREQ and RBLA do), but also takes into account
the parallelism between those requests to estimate their individual contribution
to the overall application's stall time. Therefore, UBM-ST can reduce stall time
more effectively compared with those prior proposals.  Figure~\ref{fig:Total-stall-time}
shows the sum of stall time for each workload. From this figure,
we can see that UBM-ST consistently achieves a smaller application
stall time compared with the prior mechanisms in workload categories with a non-zero memory intensity. For the most memory-intensive
category, UBM-ST achieves a 8\% stall time reduction over RBLA.  When none of
the workloads are memory intensive, UBM-ST performs on par with all prior proposals.

On the top of UBM-ST, UBM also consider the sensitivity
of system performance to the stall time of each application. 
Figure~\ref{fig:Performance-for-baseline} shows that UBM improves upon the performance
of UBM-ST for memory-intensive workloads. For the most
memory-intensive workload category, UBM gets a 7\% average performance improvement
over UBM-ST, and achieves an average performance gain of 14\% over RBLA. The maximum performance gain of UBM over RBLA is up to 39\%. For the
non-memory-intensive category, UBM achieves similar performance to
UBM-ST. This is because the sensitivity of system performance to the
stall time of different applications depends on the interference between
applications. If interference between applications is not severe,
system performance will be equally sensitive to the stall time of
each application. For the non-memory-intensive categories, since there
are fewer memory requests to be serviced, the interference between
applications is low enough so that the stall time of each application influences
the system performance by roughly equal degrees.  For the memory-intensive
categories, system performance exhibits diverse sensitivity to the
stall time of each application, and UBM takes advantage of
this diversity to further optimize system performance.

Figure~\ref{fig:Normalized-harmonic-speedup} shows the normalized
harmonic speedup of these five mechanisms. We can see that UBM-ST
and UBM consistently outperform prior proposals in all workload
categories with a non-zero memory intensity. For the most memory-intensive
workloads, UBM-ST achieves a 7\% average performance gain over RBLA, while
UBM achieves a 14\% average performance gain over RBLA. This also demonstrates
that considering both stall time reduction and sensitivity of system
performance to application's stall time can improve system performance. 

Figure~\ref{fig:Fairness-for-baseline} shows the unfairness of these
mechanisms on the baseline system configuration. We can see that both
UBM-ST and UBM achieve equivalent or improved fairness compared with all
prior proposals.

\begin{figure*}[ht]
	\centering
	
	\begin{minipage}[h]{0.32\linewidth}
		\vspace{-0.1in}
		\centering
		\includegraphics[width=0.98\linewidth]{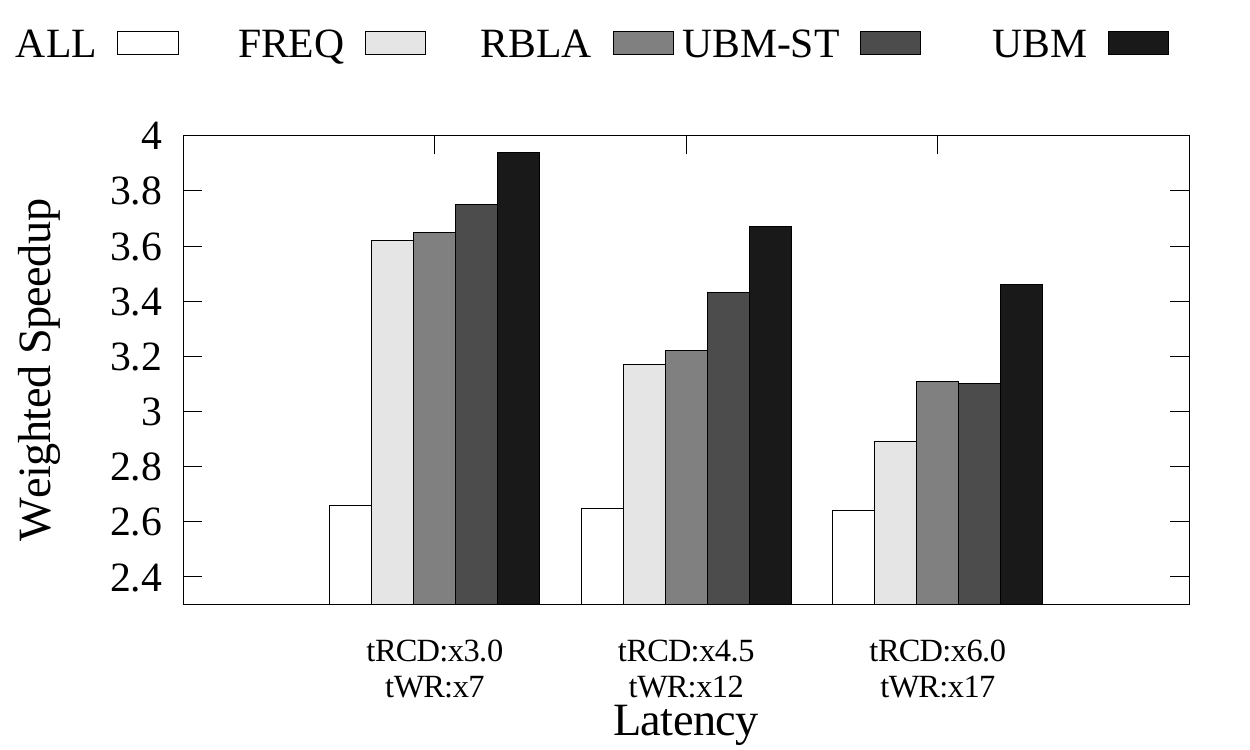}\nocaptionrule
		\caption{Weighted speedup for various NVM access latencies.}
		\label{fig:Performance-for-configurations-1}
	\end{minipage}
	\begin{minipage}[h]{0.32\linewidth}
		\vspace{-0.1in}
		\centering
		\includegraphics[width=0.98\linewidth]{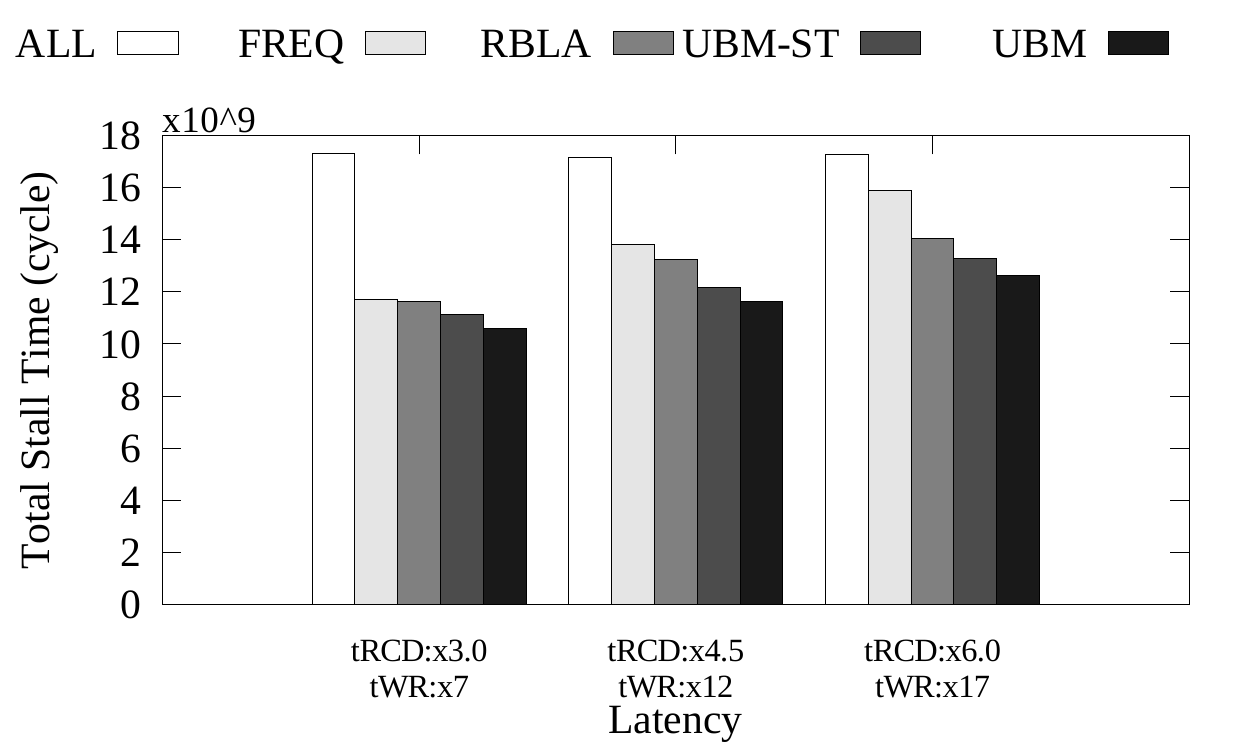}\nocaptionrule
		\caption{Total stall time for various NVM access latencies.}
		\label{fig:Total-stall-time-2}
	\end{minipage}
	\begin{minipage}[h]{0.32\linewidth}
		\vspace{-0.1in}
		\centering
		\includegraphics[width=0.98\linewidth]{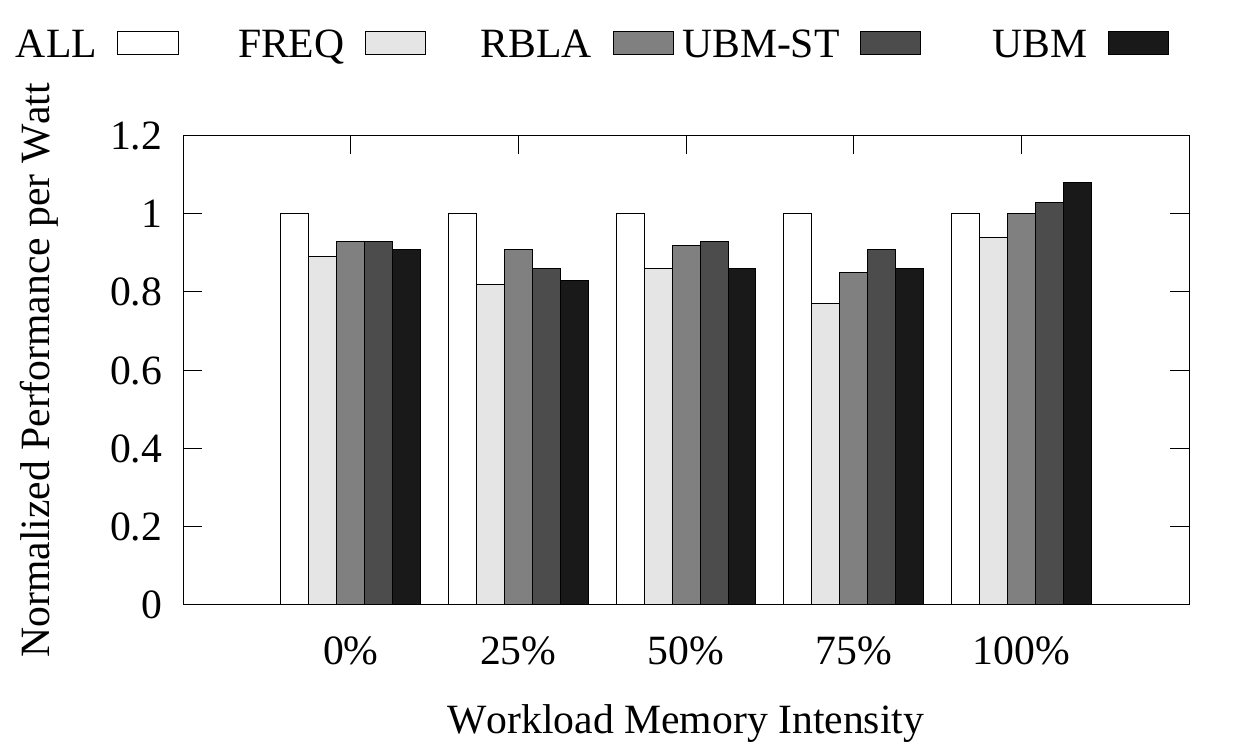}\nocaptionrule
        \caption{Energy efficiency for baseline configurations.}
        \label{fig:energy}
	\end{minipage}
\end{figure*}

\subsection{Evaluation for Various DRAM Sizes}
The DRAM size determines the room for performance optimization in hybrid
memory systems. A large DRAM can allow more pages to migrate from
NVM, offering greater system performance. However, the DRAM size cannot
realistically be too large, since DRAM is the scaling bottleneck for hybrid memory
systems. In this section, we evaluate each
mechanism for DRAM sizes of 256MB, 512MB, 1GB, and 2GB. 

Figure~\ref{fig:Performance-for-configurations} shows the weighted speedup
of workloads with 100\% memory intensity under various DRAM sizes.
We see that the system performance increases with DRAM size.
This is because a larger portion of the application working sets
can be placed in a larger DRAM. Under the four
evaluated sizes, UBM outperforms RBLA by 10\%, 14\%, 13\%, and
13\%, respectively. 
Even for a 256MB DRAM, which offers fewer opportunities for optimization,
UBM achieves a weighted speedup of 3.50,
which is larger than RBLA's weighted speedup of 3.27 \emph{for a 2GB DRAM}
(i.e., UBM can achieve RBLA's performance with only an eighth of the RAM).
This implies that by quantifying the performance benefits of each
page and selectively placing critical pages in DRAM, we can enable a shrink of
the DRAM size without performance degradation, improving memory system scalability.

Figure~\ref{fig:Total-stall-time-1} shows the sum of the stall times for
each workload. We observe that stall time decreases as DRAM size
increases. In addition, UBM achieves a stall time reduction of 11\%,
12\%, 12\%, and 12\% over RBLA, respectively, under the four DRAM sizes.

\subsection{Evaluation for Various NVM Access Latencies}

In this section, we vary the NVM access latency to test the sensitivity of
our proposed
mechanisms. In NVM, row activation time $t_{RCD}$
and write recovery time $t_{WR}$ are two important timing parameters that
influence the read/write access latency~\cite{justin_report}. $t_{RCD}$
specifies the latency between the row activate and buffer read/write
commands, while $t_{WR}$ specifies the latency between the array write
and precharge commands.

Figure~\ref{fig:Performance-for-configurations-1} shows the weighted
speedup under different NVM access latency combinations. In this figure,
$t_{RCD}$ for NVM is chosen to be 3.0, 4.5, and 6.0 times the DRAM $t_{RCD}$; $t_{WR}$ for NVM is chosen as
7, 12, and 17 times the $t_{WR}$ of DRAM. From this figure, we can
see that as $t_{RCD}$ and $t_{WR}$ increase, the system performance
gradually decreases. This is because the increased access latency
will increase the processor stall time, and in turn decrease
system throughput. The performance of ALL does not significantly change.
This is because ALL tries to insert the whole working set into DRAM,
which will lead to serious DRAM contention. Unlike the other mechanisms,
this contention, and not the NVM latency, is the bottleneck for ALL. For
the other mechanisms, since they can implicitly perform some form of load
balancing between DRAM and NVM (through the dynamic adjustment of
the migration threshold), their main bottleneck is the latency asymmetry
between DRAM and NVM, and as a result their absolute performance improves when NVM
latency decreases. For our three latency configurations, UBM achieves a weighted
speedup of 8\%, 14\%, and 11\%, respectively, over RBLA. Figure~\ref{fig:Total-stall-time-2} shows the sum of the stall times for each workload. UBM 
reduces the stall time over RBLA by 9\%, 12\%, and 10\%, respectively.

\subsection{Evaluation for Energy Efficiency}

We also study the energy efficiency of these mechanisms
on the baseline DRAM/NVM configuration. Figure~\ref{fig:energy} shows the
energy efficiency of these mechanisms on workloads with varying
memory intensities. Similar to the RBLA work~\cite{RowBufferLocality}, we use the
\textit{normalized performance per watt} metric to characterize energy
efficiency. From Figure~\ref{fig:energy}, we can see that ALL generally
has the highest energy efficiency. This is because ALL tends to insert
all its working set into DRAM, resulting in the majority of its memory
requests occuring in DRAM.
For the other four mechanisms, since they all try to
balance the bandwidth consumption between DRAM and NVM instead of placing
all of the working set into DRAM, their energy consumption is generally
higher. FREQ has lower energy efficiency compared with RBLA, UBM-ST,
and UBM. This is because FREQ does not consider row buffer locality
in its data placement decisions: As mentioned previously, row buffer
hit requests consume similar energy in DRAM and NVM, while row buffer
misses consume much higher energy in NVM, making it more energy efficient
to place pages with low row buffer hit rates in DRAM. 
Figure~\ref{fig:energy} show that RBLA, UBM-ST, and UBM all achieve
similar energy efficiency, as they all incorporate
row buffer locality into their data placement decisions. 

\section{Related Work}

To our knowledge, we provide the first utility metric for hybrid DRAM-NVM
memory systems that quantifies the system performance benefits of placing
pages in DRAM. We also provide the first comprehensive performance model for doing so. The most closely related work is a set of proposals on data placement
in hybrid memory systems. 

We have already compared our proposal (UBM) to three state-of-the-art mechanisms --- a conventional cache insertion mechanism (similar to \cite{Qureshi_2009}), an access frequency based mechanism (similar to \cite{chop, Ramos_2011}), and a row buffer locality based mechanism \cite{RowBufferLocality} --- and have shown that UBM outperforms all of them significantly (see Section~\ref{sub:EvaluationBaseline}). In this section, we discuss work in hybrid memory systems as well as other related work.

\subsection{Hybrid DRAM-NVM Memory Systems}

Qureshi et al. \cite{Qureshi_2009} propose to use DRAM as a conventional
cache for NVM to reduce its access latency.\footnote{\label{MechanismComparison}We have compared our proposed mechanism (UBM) with these (or similar) mechanisms.} Zhang and Li \cite{Zhang_3d_pact2009}
propose to mitigate the long write latency of NVM by migrating pages
that experience large numbers of write accesses to DRAM. Ramos et al.
\cite{Ramos_2011} propose to rank pages based on their access frequency
and write intensity, and migrate pages with high ranks to DRAM.\footnotemark[\value{footnote}] Yoon
et al. \cite{RowBufferLocality} observe that DRAM and NVM yield
similar latency for row buffer hit accesses but different latencies
for row buffer miss accesses, and propose to migrate pages with high
access frequency and low row buffer locality to DRAM to reduce the
average access latency.\footnotemark[\value{footnote}] These prior works only use a few
aspects of memory characterization to construct a heuristic that optimizes
access latency, not overall system performance directly. 
As previously pointed out, improving the access latency of individual
requests does not necessarily lead to a considerable improvement 
in system performance. In order to maximize system performance,
it is necessary to quantify the performance benefit of placing each
page in DRAM. Dhiman et al. \cite{PDRAM_2009} propose a method to
overcome the wear leveling issues of NVM. This method monitors the
write intensity of each page in NVM, and copies the contents of a
page to another location (either in DRAM or NVM) if its
write access count exceeds a threshold. The main goal of this method is
to alleviate the overheads associated with wear leveling, while the
main target of UBM is to improve system performance. In addition,
our proposed method is complementary with \cite{PDRAM_2009}, and can be combined to improve both
the system performance and wear leveling.

\subsection{Heterogeneous DRAM Memory Systems}

Jiang et al. \cite{chop} propose to only cache hot pages in an on-chip
DRAM cache, to overcome the off-chip DRAM bandwidth bottleneck.\footnotemark[\value{footnote}] Chatterjee
et al. \cite{critical_word} observe that the first word of cache blocks
is usually critical to the system performance, and propose to store
these words in fast DRAM. UBM is complementary to these proposals and can
be combined with both. For example, UBM can additionally store the
first words of cache blocks from pages with high utility in DRAM to
improve the system performance. Phadke and Narayanasamy \cite{mlp_heterogenous_memory}
propose to classify applications as latency-sensitive, bandwidth-sensitive,
or insensitive-to-both based on the MLP property of applications,
and run applications in DRAM with corresponding characteristics.
To estimate MLP, they use an offline approach to profile applications
in the compilation stage, measuring their MPKI and processor stall
time, and treat applications with high MPKI but low stall time as
the ones with good MLP properties. Compared with this method, the MLP
estimation approach in UBM exhibits two major differences: (1) UBM
estimates MLP using an online approach that covers the dynamic events
during program execution; (2) UBM considers the MLP effects at a page
granularity, and differentiates between pages with diverse MLP properties within
the same application.

\subsection{Other Work}
Several other works take advantage of the concepts of memory level parallelism and utility based resource management. For example, Mutlu et al. \cite{PAR-BS} propose a memory scheduler which exploits
bank level parallelism and schedules concurrent requests going to
different banks in bursts. This work is orthogonal to our technique. Qureshi et al. \cite{MLPCacheReplacement}
propose an on-chip cache replacement policy that tends to evict cache
blocks with larger MLP. The context of this work is different from
ours: it targets on-chip cache replacement, while our
work targets off-chip hybrid memory page placement. As a result, we face a more complex problem with much larger design space. For on-chip DRAM caches, retrieving data from the on-chip cache is clearly preferred over retrieving data from main memory, due to the off-chip communication latency. If it were possible, those systems would prefer that all data be kept in the on-chip cache. In contrast, both our DRAM and NVM are off-chip, with their row buffer hits having identical access latencies. Since the DRAM and NVM have separate data channels, our partitioning mechanism also performs load balancing --- as discussed in Section \ref{sub:WorkloadsSection}, some of our applications never fill the DRAM cache in order to exploit the NVM bandwidth. When combined with the fact that NVM writes are more costly than reads, the decision space for our hybrid memory becomes much more complex than that of traditional DRAM caches.

Several utility based mechanisms have also been proposed to guide cache
partitioning. Stone et al. \cite{OptimalCachePartitioning} propose
an optimal cache partitioning mechanism that uses marginal utility
(i.e., the cache hit rate gain if an application gains one more block)
to determine which application should receive the next available cache block
to maximize the overall cache hit rate. Qureshi et al. \cite{UCP} also
propose a utility-based cache partitioning mechanism which can estimate
marginal utility online. All these works are orthogonal to ours.

\section{Conclusion}

We propose a page utility based hybrid memory management mechanism
(UBM), the first mechanism to quantify the system performance benefits of
placing a page in DRAM versus NVM for hybrid memory systems. UBM comprehensively considers the interaction between
access frequency, row buffer locality and memory level parallelism
of a page to systematically estimate the stall time
reduction of placing the page in DRAM versus NVM. UBM also observes
that the system performance may exhibit different sensitivity to the
stall time of different applications, and provides a method to estimate
this sensitivity. Based on these new performance models, UBM estimates each page's
utility and migrates pages with high utility to DRAM. Experimental
results show that UBM improves the system performance by 14\% on average (and up to 39\%) over
the best of three state-of-the-art proposals. We also evaluate UBM under
various DRAM sizes and NVM latencies and observe similar benefits
under a wide variety of configurations. We conclude that the utility metric
and utility based mechanism proposed in this paper enables an effective approach to hybrid memory management. We also hope the new utility metric introduced in this paper can be useful in solving other page migration and memory management problems.

\bibliographystyle{IEEEtranS} \bibliography{template}

\begin{thebibliography}{10}
\providecommand{\url}[1]{#1}
\csname url@samestyle\endcsname
\providecommand{\newblock}{\relax}
\providecommand{\bibinfo}[2]{#2}
\providecommand{\BIBentrySTDinterwordspacing}{\spaceskip=0pt\relax}
\providecommand{\BIBentryALTinterwordstretchfactor}{4}
\providecommand{\BIBentryALTinterwordspacing}{\spaceskip=\fontdimen2\font plus
\BIBentryALTinterwordstretchfactor\fontdimen3\font minus
  \fontdimen4\font\relax}
\providecommand{\BIBforeignlanguage}[2]{{%
\expandafter\ifx\csname l@#1\endcsname\relax
\typeout{** WARNING: IEEEtranS.bst: No hyphenation pattern has been}%
\typeout{** loaded for the language `#1'. Using the pattern for}%
\typeout{** the default language instead.}%
\else
\language=\csname l@#1\endcsname
\fi
#2}}
\providecommand{\BIBdecl}{\relax}
\BIBdecl

\bibitem{itrs}
``Process integration, devices, and structures,'' in \emph{The International
  Technology Roadmap for Semiconductors}, 2013.

\bibitem{maxslowdown}
M.~A. Bender, S.~Chakrabarti, and S.~Muthukrishnan, ``Flow and stretch metrics
  for scheduling continuous job streams,'' in \emph{SODA}, 1998.

\bibitem{critical_word}
N.~Chatterjee \emph{et~al.}, ``Leveraging heterogeneity in {DRAM} main memories
  to accelerate critical word access,'' in \emph{MICRO}, 2012.

\bibitem{JSSC2013}
K.~C. Chun \emph{et~al.}, ``A scaling roadmap and performance evaluation of
  in-plane and perpendicular {MTJ} based {STT-MRAMs} for high-density cache
  memory,'' \emph{JSSC}, vol.~48, no.~2, Feb 2013.

\bibitem{IEDM2010}
S.~Chung \emph{et~al.}, ``Fully integrated 54nm {STT-RAM} with the smallest bit
  cell dimension for high density memory application,'' in \emph{IEDM}, 2010.

\bibitem{YCSB}
B.~F. Cooper \emph{et~al.}, ``Benchmarking cloud serving systems with {YCSB},''
  in \emph{SOCC}, 2010.

\bibitem{PDRAM_2009}
G.~Dhiman, R.~Ayoub, and T.~Rosing, ``{PDRAM}: A hybrid {PRAM} and {DRAM} main
  memory system,'' in \emph{DAC}, 2009.

\bibitem{metrics}
S.~Eyerman and L.~Eeckhout, ``System-level performance metrics for multiprogram
  workloads,'' \emph{IEEE Micro}, vol.~28, no.~3, May 2008.

\bibitem{Ghose2013}
S.~Ghose, H.~Lee, and J.~F. Mart\'{\i}nez, ``Improving memory scheduling via
  processor-side load criticality information,'' in \emph{ISCA}, 2013.

\bibitem{SpecCpu}
J.~L. Henning, ``{SPEC CPU2006} benchmark descriptions,'' \emph{SIGARCH Comput.
  Archit. News}, vol.~34, no.~4, Sep. 2006.

\bibitem{chop}
X.~Jiang \emph{et~al.}, ``{CHOP}: Adaptive filter-based {DRAM} caching for
  {CMP} server platforms,'' in \emph{HPCA}, 2010.

\bibitem{clear}
N.~Kirman \emph{et~al.}, ``Checkpointed early load retirement,'' in
  \emph{HPCA}, 2005.

\bibitem{Architect_STTRAM}
E.~Kultursay \emph{et~al.}, ``Evaluating {STT-RAM} as an energy-efficient main
  memory alternative,'' in \emph{ISPASS}, 2013.

\bibitem{Architect_PCM_ISCA2009}
B.~C. Lee \emph{et~al.}, ``Architecting phase change memory as a scalable
  {DRAM} alternative,'' in \emph{ISCA}, 2009.

\bibitem{Architect_PCM_MicroTopPicks}
B.~C. Lee \emph{et~al.}, ``Phase-change technology and the future of main
  memory,'' \emph{IEEE Micro}, vol.~30, no.~1, Jan 2010.

\bibitem{ReRam}
T.~Liu \emph{et~al.}, ``A 130.7mm2 2 layer 32{Gb} {ReRAM} memory device in 24nm
  technology,'' \emph{JSSC}, vol.~49, no.~1, Jan 2014.

\bibitem{pin}
C.-K. Luk \emph{et~al.}, ``{Pin}: Building customized program analysis tools
  with dynamic instrumentation,'' in \emph{PLDI}, 2005.

\bibitem{harmonic_speedup}
K.~Luo, J.~Gummaraju, and M.~Franklin, ``Balancing thoughput and fairness in
  {SMT} processors,'' in \emph{ISPASS}, 2001.

\bibitem{lbm_2002}
J.~Mandelman \emph{et~al.}, ``Challenges and future directions for the scaling
  of dynamic random-access memory ({DRAM}),'' \emph{IBM J. Res. Dev.}, vol.~46,
  no. 2.3, March 2002.

\bibitem{Meza_2012_cal}
J.~Meza \emph{et~al.}, ``Enabling efficient and scalable hybrid memories using
  fine-granularity {DRAM} cache management,'' \emph{IEEE CAL}, Jul. 2012.

\bibitem{justin_report}
J.~Meza, J.~Li, and O.~Mutlu, ``Evaluating row buffer locality in future
  non-volatile main memories,'' in \emph{SAFARI Technical Report}, 2012.

\bibitem{Micron}
Micron, ``1{G}b: x4, x8, x16 {DDR3 SDRAM},'' 2013.

\bibitem{STF_StallTimeFair}
O.~Mutlu and T.~Moscibroda, ``Stall-time fair memory access scheduling for chip
  multiprocessors,'' in \emph{MICRO}, 2007.

\bibitem{runahead}
O.~Mutlu \emph{et~al.}, ``Runahead execution: an alternative to very large
  instruction windows for out-of-order processors,'' in \emph{HPCA}, 2003.

\bibitem{PAR-BS}
O.~Mutlu and T.~Moscibroda, ``Parallelism-aware batch scheduling: Enhancing
  both performance and fairness of shared dram systems,'' in \emph{ISCA}, 2008.

\bibitem{mlp_heterogenous_memory}
S.~Phadke and S.~Narayanasamy, ``{MLP} aware heterogeneous memory system,'' in
  \emph{DATE}, 2011.

\bibitem{UCP}
M.~Qureshi and Y.~Patt, ``Utility-based cache partitioning: A low-overhead,
  high-performance, runtime mechanism to partition shared caches,'' in
  \emph{MICRO}, 2006.

\bibitem{MLPCacheReplacement}
M.~K. Qureshi \emph{et~al.}, ``A case for mlp-aware cache replacement,'' in
  \emph{ISCA}, 2006.

\bibitem{Qureshi_2009}
M.~K. Qureshi, V.~Srinivasan, and J.~A. Rivers, ``Scalable high performance
  main memory system using phase-change memory technology,'' in \emph{ISCA},
  2009.

\bibitem{Ramos_2011}
L.~E. Ramos, E.~Gorbatov, and R.~Bianchini, ``Page placement in hybrid memory
  systems,'' in \emph{ICS}, 2011.

\bibitem{ibm_2008}
S.~Raoux \emph{et~al.}, ``Phase-change random access memory: A scalable
  technology,'' \emph{IBM J. Res. Dev.}, vol.~52, no.~4, Jul. 2008.

\bibitem{FRFCFS1}
S.~Rixner \emph{et~al.}, ``Memory access scheduling,'' in \emph{ISCA}, 2000.

\bibitem{WeightedSpeedUp}
A.~Snavely and D.~M. Tullsen, ``Symbiotic jobscheduling for a simultaneous
  multithreaded processor,'' in \emph{ASPLOS}, 2000.

\bibitem{OptimalCachePartitioning}
H.~S. Stone, J.~Turek, and J.~Wolf, ``Optimal partitioning of cache memory,''
  \emph{Computers, IEEE Transactions on}, 1992.

\bibitem{IEDM2009}
Y.-H. Tseng \emph{et~al.}, ``High density and ultra small cell size of contact
  {ReRAM (CR-RAM)} in 90nm {CMOS} logic technology and circuits,'' in
  \emph{IEDM}, 2009.

\bibitem{RowBufferLocality}
H.~Yoon \emph{et~al.}, ``Row buffer locality aware caching policies for hybrid
  memories,'' in \emph{ICCD}, 2012.

\bibitem{Zhang_3d_pact2009}
W.~Zhang and T.~Li, ``Exploring phase change memory and {3D} die-stacking for
  power/thermal friendly, fast and durable memory architectures,'' in
  \emph{PACT}, 2009.

\bibitem{ReadWriteRatio}
M.~Zhou \emph{et~al.}, ``Writeback-aware bandwidth partitioning for multi-core
  systems with {PCM},'' in \emph{PACT}, 2013.

\bibitem{ISCA2009_Pittsburgh}
P.~Zhou \emph{et~al.}, ``A durable and energy efficient main memory using phase
  change memory technology,'' in \emph{ISCA}, 2009.

\bibitem{FRFCFS2}
W.~Zuravleff and T.~Robinson, ``Controllers for a synchronous {DRAM} that
  maximizes throughput by allowing memory requests and commands to be issued
  out of order,'' in \emph{U.S. Patent Number 5,630,096}, 1997.

\end{thebibliography}
\end{document}